\documentclass[aps,prx,superscriptaddress,amsfonts,amsmath,amssymb,showpacs,floatfix,reprint]{revtex4-1}

\usepackage[english]{babel}
\usepackage{graphicx}					
\usepackage[pdftex]{epsfig}
\usepackage{ragged2e}
\usepackage[caption=false]{subfig}
\usepackage{amsmath,amssymb}
\usepackage{bbm}
\usepackage{bm}
\usepackage{times}
\usepackage[colorlinks,linkcolor=blue,citecolor=blue,urlcolor=blue]{hyperref}
\usepackage{color}
\usepackage[table,xcdraw]{xcolor}
\usepackage{epstopdf}
\usepackage{physics}
\usepackage[export]{adjustbox}
\usepackage{dsfont}
\usepackage{float}
\usepackage{makecell}
\usepackage{subfloat}

\newcommand{\be}{\begin{equation}}
\newcommand{\ee}{\end{equation}}

\newcommand{\1}{\hspace*{-1pt}}
\newcommand{\2}{\hspace*{-2pt}}

\newcommand{\bee}{\begin{equation*}}
\newcommand{\eee}{\end{equation*}}

\renewcommand\[{\begin{equation}}
\renewcommand\]{\end{equation}}

\definecolor{cadmiumgreen}{rgb}{0.0, 0.42, 0.24}

\begin{document}

\title{Fracton excitations in classical frustrated kagome spin models}

\author{Max Hering}
\affiliation{Helmholtz-Zentrum Berlin f\"{u}r Materialien und Energie, Hahn-Meitner Platz 1, 14109 Berlin, Germany}
\affiliation{Dahlem Center for Complex Quantum Systems and Fachbereich Physik, Freie Universit\"at Berlin, 14195 Berlin, Germany}
\author{Han Yan}
\affiliation{Theory of Quantum Matter Unit, Okinawa Institute of Science and Technology Graduate University, Onna-son, Okinawa 904-0412, Japan}
\author{Johannes Reuther}
\affiliation{Helmholtz-Zentrum Berlin f\"{u}r Materialien und Energie, Hahn-Meitner Platz 1, 14109 Berlin, Germany} 
\affiliation{Dahlem Center for Complex Quantum Systems and Fachbereich Physik, Freie Universit\"at Berlin, 14195 Berlin, Germany}

\date{\today}

\begin{abstract}
Fractons are topological quasiparticles with   limited mobility. While there exists a variety of models hosting these excitations, typical fracton systems require rather complicated many-particle interactions. Here, we discuss fracton behavior in the more common physical setting of classical kagome spin models with frustrated two-body interactions only. We investigate systems with different types of elementary spin degrees of freedom (three-state Potts, XY, and Heisenberg spins) which all exhibit   characteristic subsystem symmetries and fracton-like excitations. The mobility constraints of isolated fractons and bound fracton pairs in the three-state Potts model are, however, strikingly different compared to the known type-I or type-II fracton models. One may still explain these properties in terms of type-I fracton behavior and construct an effective low-energy tensor gauge theory when considering the system as a 2D cut of a 3D cubic lattice model. Our extensive classical Monte-Carlo simulations further indicate a crossover into a low temperature glassy phase where the system gets trapped in metastable fracton states. Moving on to XY spins, we find that in addition to fractons the system hosts fractional vortex excitations. As a result of the restricted mobility of both types of defects, our classical Monte-Carlo simulations do not indicate a Kosterlitz-Thouless transition but again show a crossover into a glassy low-temperature regime. Finally, the energy barriers associated with fractons vanish in the case of Heisenberg spins, such that defect states may continuously decay into a ground state. These decays, however, exhibit a power-law relaxation behavior which leads to slow equilibration dynamics at low temperatures.
\end{abstract}

\maketitle

\section{Introduction}
Fractional quasiparticles are a widespread phenomenon in currently investigated condensed matter phases. Despite their rather conventional constituents such as the electron's charge or spin, the actual low-energy physics is governed by excitations that are fractions of the original degrees of freedom. Typical examples are fractional quantum Hall systems~\cite{tsui82,laughlin83,halperin84} or quantum spin liquids~\cite{balents10,broholm20,savary16,zhou17} where the occurrence of fractional quasiparticles is closely tied to topological order and long-range entangled ground states~\cite{wen91}. While the exotic nature of fractional excitations manifests in various intriguing ways (e.g. anyonic braiding statistics), they usually still possess the very common property of being equipped with a kinetic degree of freedom.

In the last few years, however, a new class of systems has attracted increasing interest, where the low-energy fractional excitations are intrinsically {\it immobile}, known as fractons~\cite{pretko20,nandkishore19,chamon05,haah11,vijay15,vijay16,castelnovo10,yoshida13,bravyi11,shirley18,shirley19,han19,han19_2}. In simple terms, the immobility stems from the fact that fractons can only be created at the corners of a membrane-like operator, known as {\it type-I} scenario~\cite{vijay16}. This is, e.g., in contrast to a more conventional quantum spin liquid, where fractional spin excitations appear at the ends of a string operator~\cite{savary16}. Stated differently, in these latter systems the fractional `charge' is conserved and the dipole operator is unconserved while a fracton phase is characterized by both conserved charge {\it and} dipole moments. Such dipole conservation laws are a natural property of symmetric tensor gauge theories which provide an effective low-energy description of fracton phases~\cite{pretko17,pretko17_2,pretko18,bulmash18,bulmash18_2,ma18}. Despite their immobility when being isolated, composites of fractons may be partially mobile within lower dimensional subsystems. Such subsystem operations establish a subextensive groundstate degeneracy characteristic for many fracton phases.

As a result of their mobility constraints, fracton phases feature a slow thermal relaxation and glassy dynamics even in the absence of randomness in the Hamiltonian~\cite{chamon05,prem17}. While in disorder-induced spin glasses the slow dynamics is a consequence of a complex distribution of energy barriers, in fracton phases the glassy behavior rather stems from restrictions in the elementary moves by which the system can transition between states. Fracton phases, hence, share many properties with kinetically constrained models~\cite{ritort03,newman99,garrahan00,jack05,lipowski97,garrahan02,espriu04,keys11,frederickson84}, which have a much longer history of investigation.

A severe difficulty in realizing fracton phases in real physical systems is that most of the known models consist of rather artificial and complicated cluster spin interactions. For example, the famous X-cube model has interactions involving twelve spin operators~\cite{vijay16}. Various interesting proposals have recently been put forward, aiming to embed fracton physics in real world systems, however, this field is still in its infancy. For example, it has been shown that in a certain limit of coupled Kitaev honeycomb layers, type-I fracton order emerges with nearest-neighbor two-body spin interactions only~\cite{slagle17}. In other works, fracton mobility constraints have been identified in valence plaquette solids~\cite{you20}, frustrated hole-doped antiferromagnets~\cite{sous20},
and breathing pyrochlore magnets~\cite{YanPRL2020}.

Here, we investigate fracton behavior in the familiar context of frustrated two-body spin models on the kagome lattice. Our conceptual starting point is a classical kagome spin model with nearest neighbor interactions~\cite{reimers93,chalker92,ritchey93,zhitomirksy08}, featuring an extensive ground state degeneracy. Through a suitable inclusion of longer-range couplings, this degeneracy is lifted to become subextensive, hence, realizing an environment for fracton physics (see Refs.~\cite{messio11,messio12,bernu13,gomez18,iida20,mizoguchi17,mizoguchi18,suttner14,iqbal15,balents02,rehn17} for a selection of works about related kagome models with longer-range interactions). The ground and excited states may be most conveniently described by defect variables associated with local spin constraints. We study three variants of this system where the elementary constituents become increasingly more realistic: Starting with a three-state Potts model we generalize the spins towards continuous inplane XY degrees of freedom and finally consider isotropic Heisenberg spins. While all our models are classical spin systems which cannot display real fracton topological order, we still identify characteristic fracton phenomena such as subdimensional particles, emergent tensor gauge theories and glassy dynamics.

A first observation is that our three-state kagome Potts model hosts isolated defects which correspond to a single violated spin constraint. While the restricted mobility of such excitations shows a strong resemblance with fractons, they still do not fall exactly into the known type-I~\cite{vijay16} or type-II~\cite{haah11} fracton categories. Particularly, defects are neither created at the corners of a modified region nor through a fractal operation. We resolve this mystery by embedding the kagome structure in a simple 3D  cubic lattice. This allows us to view the kagome Potts model as a more conventional 3D type-I fracton system, restricted to a particular two-dimensional subspace. We further formulate a 3D rank-2 U(1) electrostatics theory describing the system's low-energy behavior. Concerning thermal properties, our extensive classical Monte Carlo simulations indicate a high temperature regime where the system shows characteristic line-like spin fluctuations on short length scales. In contrast, at low temperatures the system enters a glassy regime where the dynamics slows down and spin configurations get stuck in (or around) local energy minima (see e.g. Refs.~\cite{cepas12,bilitewski17,bilitewski19,hamp18,ferrero03,bekhechi03} for further works on glassy dynamics in kagome spin systems).

Generalizing the spins towards XY degrees of freedom, the isolated fractons remain qualitatively unchanged. However, the increased configurational space enables the existence of vortices with fractional vorticity, known from other classical XY kagome models~\cite{rzchowski97,korshunov02}. Our Monte Carlo simulations reveal thermally excited patterns of fractons and fractional vortices whose positions are strongly correlated among each other. These correlations significantly reduce the dynamics of vortices such that a Kosterlitz-Thouless transition~\cite{kosterlitz73,berezinskii71,berezinskii72} into a quasi long-range ordered low-temperature phase is not observed. Finally, in the case of Heisenberg spins, any defect state can be continuously transformed into a ground state without crossing energy barriers and, consequently, fractons lose their stability. The associated time scales, however, easily exceed available computation times such that even in slowly cooled systems remnants of fracton states are still discernible.

In total, this work demonstrates that fracton behaviors are not restricted to models with artificial spin cluster interactions but may be observed in more common frustrated two-body spin systems.

The remainder of this work is organized as follows: In Sec.~\ref{sec:models} we introduce the investigated models which are defined in terms of three-color states, XY spins, and Heisenberg spins. In the following Sec.~\ref{sec:analytic} we discuss the properties of ground states and isolated defect states for all three variants of the system. We, particularly, focus on fractons in the three-state Potts model (Sec.~\ref{sec:analytic_potts}) and explain their low energy behaviors and effective field theory. Thereafter, in Sec.~\ref{sec:numerics} we investigate the thermal properties of the three systems using classical Monte Carlo simulations. We discuss in detail thermodynamic quantities such as specific heat, spin-structure factor, real space spin configurations, as well as autocorrelation functions. The paper ends with a conclusion in Sec.~\ref{sec:conclusion}.

\section{Models}\label{sec:models}
\begin{figure}
\includegraphics[width=0.8\linewidth]{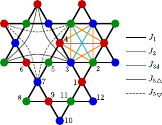}
\caption{Definition of different bonds on the kagome lattice. The hexagon on the top left illustrates the fifths neighbor $J_{5\bigtriangleup}$ and $J_{5\bigtriangledown}$ bonds while the hexagon on the top right shows examples for $J_2$ and $J_{3\text{d}}$-bonds. The numbers $1-12$ label a set of twelve spins forming a Star of David, see text for details. The illustrated spin state is an example for a configuration fulfilling all constraints discussed in the main text.\label{fig:models}}
\end{figure}

Below we study a family of classical three-state Potts, XY and Heisenberg spin models defined on the sites of the kagome lattice. Depending on the type of model, the spin operators ${\mathbf S}_i$, hence, either denote color states ${\mathbf S}_i\in\{\text{red, blue, green}\}$ [or equivalently ${\mathbf S}_\text{red}=(1,0,0)$, ${\mathbf S}_\text{blue}=(-1/2,\sqrt{3}/2,0)$, ${\mathbf S}_\text{green}=(-1/2,-\sqrt{3}/2,0)$], XY spins ${\mathbf S}_i=(S_i^x,S_i^y,0)$, or Heisenberg spins ${\mathbf S}_i=(S_i^x,S_i^y,S_i^z)$ where normalization $|{\mathbf S}_i|=1$ is always assumed. The ground states of all models are subject to two types of constraints: Each elementary nearest neighbor triangle and six-site hexagon of the kagome lattice is locally in a ground state if their spins sum up to zero (also referred to as color-neutrality in the Potts model), 
\begin{equation}
\text{$J_1$-$J_2$-$J_{3\text{d}}$ model:}\sum_{i\in\text{n.n. triangle}}{\mathbf S}_i=0\;,\sum_{i\in\text{hexagon}}{\mathbf S}_i=0\;.
\end{equation}
In Fig.~\ref{fig:models} we present a possible three-color configuration whose construction will be discussed in the next section. We call this model the $J_1$-$J_2$-$J_{3\text{d}}$ model for reasons to become clear below. Triangles and hexagons violating this rule are associated with an energy cost.

Furthermore, we construct two variants of this model which share the same ground states but partially differ in their excited states. To this end, consider the twelve spins labelled ${\mathbf S}_1,\ldots,{\mathbf S}_{12}$ in Fig.~\ref{fig:models} which form a Star of David. Combining the constraints for three triangles with the one for the hexagon leads to
\begin{align}
&0\1=\1({\mathbf S}_1+{\mathbf S}_{11}+{\mathbf S}_{12})+({\mathbf S}_3+{\mathbf S}_{4}+{\mathbf S}_{5})+({\mathbf S}_7+{\mathbf S}_{8}+{\mathbf S}_{9})\;\notag\\
&=\underbrace{({\mathbf S}_1+{\mathbf S}_{3}+{\mathbf S}_{5}+{\mathbf S}_7+{\mathbf S}_{9}+{\mathbf S}_{11})}_{=0}+({\mathbf S}_4+{\mathbf S}_{8}+{\mathbf S}_{12})\;,\label{equivalence}
\end{align}
indicating that the hexagon condition implies color neutrality in larger upward pointing triangles (here, the triangle formed by ${\mathbf S}_4$, ${\mathbf S}_{8}$, and ${\mathbf S}_{12}$). These triangles have a side length of three nearest neighbor lattice spacings and will be denoted as $J_{5\bigtriangleup}$ triangles since the involved spins are fifth neighbors. Hence, we may formulate a variant of the above $J_1$-$J_2$-$J_{3\text{d}}$ model which we call the $J_1$-$J_{5\bigtriangleup}$-model:
\begin{equation}
\text{$J_1$-$J_{5\bigtriangleup}$ model:}\sum_{i\in\text{n.n. triangle}}{\mathbf S}_i=0\;,\sum_{i\in J_{5\bigtriangleup}\text{ triangle}}{\mathbf S}_i=0\;.
\end{equation}
Finally, the steps in Eq.~(\ref{equivalence}) can be modified to show that the color-neutrality constraint for the hexagon is also equivalent to ${\mathbf S}_2+{\mathbf S}_{6}+{\mathbf S}_{10}=0$, which involves the spins on a downward pointing large triangle (denoted as $J_{5\bigtriangledown}$ triangle). Adding this constraint results in the so-called $J_1$-$J_{5\bigtriangleup}$-$J_{5\bigtriangledown}$-model:
\begin{align}
&\text{$J_1$-$J_{5\bigtriangleup}$-$J_{5\bigtriangledown}$ model:}\notag\\
&\sum_{i\in\text{n.n. triangle}}{\mathbf S}_i=0\;,\sum_{i\in J_{5\bigtriangleup}\text{ triangle}}{\mathbf S}_i=0\;,\sum_{i\in J_{5\bigtriangledown}\text{ triangle}}{\mathbf S}_i=0\;.
\end{align}
From their construction, it is clear that all three models have the same ground states, however, their excited states differ to some extent.

The above constraints can be straightforwardly recast into quadratic spin Hamiltonians which fixes the energies of excited states. This amounts to replacing a constraint ${\mathbf S}_{i_1}+{\mathbf S}_{i_2}+\ldots+{\mathbf S}_{i_n}=0$ by a term $\sim J({\mathbf S}_{i_1}+{\mathbf S}_{i_2}+\ldots+{\mathbf S}_{i_n})^2$ in the Hamiltonian. Up to constants the above models then read
\begin{equation}
H_{\text{$J_1$-$J_2$-$J_{3\text{d}}$}}=J_1\2\sum_{\langle i,j\rangle_1}{\mathbf S}_i\cdot{\mathbf S}_{j}+J_2\2\sum_{\langle i,j\rangle_2}{\mathbf S}_i\cdot{\mathbf S}_{j}+J_{3\text{d}}\2\sum_{\langle i,j\rangle_{3\text{d}}}{\mathbf S}_i\cdot{\mathbf S}_{j}\;,
\end{equation}
where $J_2=J_{3\text{d}}$. Furthermore, since the first neighbor coupling $J_1$ has contributions from both the triangle and hexagon constraints, this interaction is bounded by $J_2=J_{3\text{d}}<J_1$. Note that $\langle i,j\rangle_x$ stands for a pair of sites coupled by $J_x$ and each pair appears in the sum once, see Fig.~\ref{fig:models} for the definition of coupling constants. Equivalently, the other models read
\begin{align}
&H_{\text{$J_1$-$J_{5\bigtriangleup}$}}=J_1\sum_{\langle i,j\rangle_1}{\mathbf S}_i\cdot{\mathbf S}_{j}+J_{5\bigtriangleup}\sum_{\langle i,j\rangle_{5\bigtriangleup}}{\mathbf S}_i\cdot{\mathbf S}_{j}\;,\\
&H_{\text{$J_1$-$J_{5\bigtriangleup}$-$J_{5\bigtriangledown}$}}=H_{\text{$J_1$-$J_{5\bigtriangleup}$}}+J_{5\bigtriangledown}\sum_{\langle i,j\rangle_{5\bigtriangledown}}{\mathbf S}_i\cdot{\mathbf S}_{j}\;.
\end{align}
As long as all interactions $J$ are positive, the properties of the three models do not crucially depend on the precise coupling ratios $\frac{J_2}{J_1}$, $\frac{J_{5\bigtriangleup}}{J_1}$ and $\frac{J_{5\bigtriangledown}}{J_1}$. Hence, without loss of generality, we will fix these ratios in our numerical calculations below. Throughout the paper, we use regular periodic boundary conditions or open boundary conditions, and do not consider twisted boundary conditions.

\section{Ground states and low-energy excitations} \label{sec:analytic}
In this section we discuss the set of ground states and isolated low-energy excitations (such as fracton-like defect states) of the models introduced in the last section, starting with the three-state Potts models. In the next section we will numerically simulate their thermodynamic properties within Monte Carlo. 

\subsection{Three-state Potts models}
\label{sec:analytic_potts}
\begin{figure}
\includegraphics[width=0.99\linewidth]{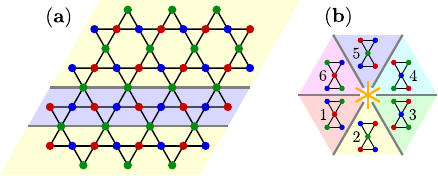}
\caption{(a) Example of a ground state of the three-state kagome Potts models. Colors have been swapped along a horizontal line, creating a stripe-like domain. (b) Adjacency rules for the six types of ${\mathbf q}=0$ orders. Neighboring and opposite sectors may share a common domain wall, see text for details.\label{fig:potts_ground_states}}
\end{figure}

\subsubsection{Ground states}
The simplest ground states of the Potts models obeying all constraints are homogeneous ${\mathbf q}=0$-states where all unit cells (which can be chosen as the three sites of the upward-pointing $J_1$ triangles) are identical. Globally, there are $3!=6$ such states which transform into each other by permutations of the three colors [see yellow region in Fig.~\ref{fig:potts_ground_states}(a) for an example of a ${\mathbf q}=0$-state]. The homogeneous ${\mathbf q}=0$-states are characterized by an alternation of two colors along straight lines running through the entire system. Starting from these ${\mathbf q}=0$-states, all other ground states can be obtained by swapping the colors along arbitrary {\it parallel} lines, which keeps all constraints intact. Note, however, that swapping the colors along two non-parallel lines creates defect triangles/hexagons near the point where the lines cross. It follows that the system has a {\it subextensive} ground state degeneracy proportional to $2^L$ (where $L$ is the linear system size) typical for fracton systems. In Fig.~\ref{fig:potts_ground_states}(a) we depict an example of a ground state where colors along a single horizontal line have been swapped with respect to the rest of the system. Note that when relaxing the constraints in the hexagons or $J_5$ triangles the ground state degeneracy is lifted, leading to the well-known extensive degeneracy characteristic for nearest neighbor kagome antiferromagnets~\cite{chalker92,reimers93}.

An alternative way of describing the states of the Potts models is by viewing the six types of ${\mathbf q}=0$-states as {\it domains}. It is then straightforward to formulate rules for possible arrangements of domains such that all constraints are respected at the domain walls. These rules are summarized in Fig.~\ref{fig:potts_ground_states}(b): Two domains in adjacent sectors can have a common domain wall with an orientation given by the gray line separating them. Additionally, domains in opposite sectors (connected by orange lines) may have a common domain wall with an orientation perpendicular to the respective orange line. Trivially, the rules remain valid when interchanging each sector in Fig.~\ref{fig:potts_ground_states}(b) with the opposite one, e.g., the domain $6$ may lie above a domain $1$ or vice versa. As illustrated in Fig.~\ref{fig:potts_ground_states}(b) domain walls respecting all constraints are straight lines passing through kagome sites of the same sublattice. Below, we will use this domain-wall representation for discussing defect states.

\subsubsection{Fracton behaviors}
We now investigate three-color states with isolated defects in single hexagons/triangles excited from a ground state. A first indication for possible fracton behavior comes from the observation that changing the value of one spin in a ground state leads to a multipole of defects, each representing a ``fraction'' of the excitation. For example, changing a single spin in a ground state of the $J_1$-$J_2$-$J_{3\text{d}}$ model creates two defect triangles and two defect hexagons adjacent to the modified spin. In the usual type-I fracton scenario, as for example realized in the plaquette Ising model (see below), these quadrupolar defects can be far-separated such that they form the corners of an area of flipped spins. Hence, the two relevant questions discussed below are: $(i)$ Do isolated defect triangles/hexagons exist in our kagome Potts model and $(ii)$ if yes, are they located at the corners of a region of flipped spins? While the answer to the first question is {\it yes}, the situation in the second question cannot be realized.
\begin{figure}
\includegraphics[width=0.8\linewidth]{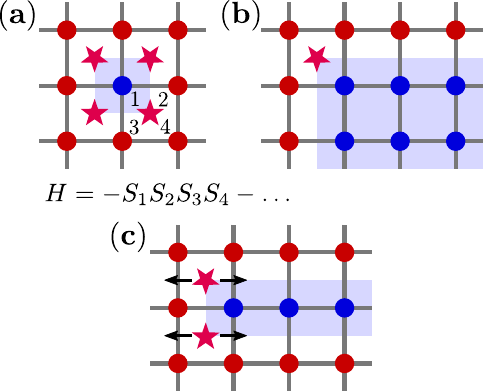}
\caption{Properties of the plaquette Ising model. (a) Flipping a single spin in a ground state creates four defect plaquettes (indicated by magenta stars). (b) When the region of flipped spins is enlarged, isolated and immobile fracton excitations sit at its corners. (c) Two-fracton bound states are free to move in the direction perpendicular to itself. \label{fig:plaquette_ising}}
\end{figure}

Before we discuss the kagome Potts models, we briefly introduce the well-known plaquette Ising model~\cite{jack05,lipowski97,garrahan02}. Despite its simplicity, this model features various prototypical type-I fracton properties which we will compare with the fracton behaviors in our kagome systems. The plaquette Ising model exhibits Ising spins $S_i=\pm1$ located at the sites of a 2D square lattice, coupled via four-body interactions $S_1 S_2 S_3 S_4$ involving the four spins of an elementary $1\times 1$ square plaquette. The Hamiltonian is a sum over all plaquette terms
\begin{equation}
H_\text{plaq.-Ising}=-\sum_\text{plaquettes}\prod_{i\in\text{plaquette}}S_i\;.
\end{equation}
Starting with a homogeneous ground state where all spins are $S_i=+1$ (or equivalently $S_i=-1$) the system's subextensive ground state degeneracy is obvious from the fact that arbitrary lines of spins (which may also intersect) can be flipped. A single spin flip at site $i$ in an arbitrary ground state creates a quadrupole of four excitations in the plaquettes sharing the site $i$, see Fig.~\ref{fig:plaquette_ising}(a). These excitations can be separated by enlarging the region of flipped spins into a rectangular area [Fig.~\ref{fig:plaquette_ising}(b)]. The isolated defects at the corners of this region, called fractons, are characterized by their immobility, since any move to a neighboring site requires flipping a whole line of spins. Despite the immobility of single fractons, however, a two-fracton bound state (so-called {\it lineon}) is free to move in the direction perpendicular to itself, as illustrated in Fig.~\ref{fig:plaquette_ising}(c). Such moves only require flipping a finite number of spins, given by the size of the dipole.
\begin{figure}
\includegraphics[width=0.95\linewidth]{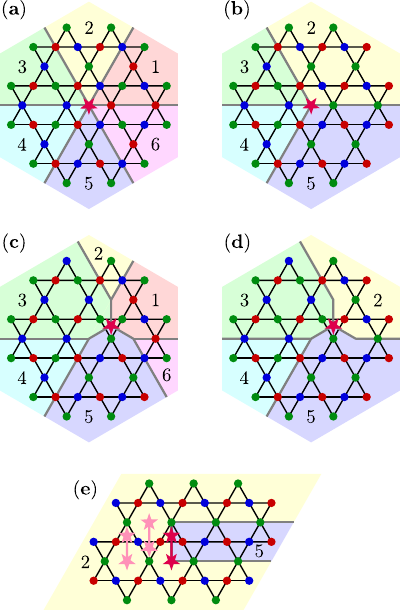}
\caption{(a)-(d) Isolated fractons (magenta stars) in the kagome Potts models residing at intersections of domain walls (gray lines). (a) Triple defect in a hexagon/$J_5$ triangle. (b) Single defect in a hexagon/$J_5$ triangle. (c) Triple defect in a $J_1$ triangle. (d) Single defect in a $J_1$ triangle. (e) Two-fracton bound state in the $J_1$-$J_2$-$J_{3\text{d}}$ model with horizontal mobility. Light magenta stars indicate the positions of fractons when the domain ``5'' (light blue region) is further extended to the left. \label{fig:potts_fractons}}
\end{figure}

We now contrast these properties with the fracton behaviors in our kagome Potts models. Firstly, isolated defect triangles/hexagons in an otherwise defect-free system exist and can be most easily described as points where domain walls cross. There are two types of domain wall arrangements which yield an excitation in a hexagon (or in a $J_{5\bigtriangleup}$/$J_{5\bigtriangledown}$ triangle): The first [called triple fracton, see Fig.~\ref{fig:potts_fractons}(a)] exhibits $\pi/3$-sectors of domains, where six domain walls emanate from the crossing point. This fracton has an excitation energy of $E_{\text{3-frac}}^{\text{Potts}}=4.5J_2$ [$4.5J_{5\bigtriangleup}$, $4.5(J_{5\bigtriangleup}+J_{5\bigtriangledown})$] for the $J_1$-$J_2$-$J_{3\text{d}}$ model [$J_1$-$J_{5\bigtriangleup}$-model, $J_1$-$J_{5\bigtriangleup}$-$J_{5\bigtriangledown}$-model]. The second [called single fracton, see Fig.~\ref{fig:potts_fractons}(b)] has two $\pi/3$ and two $2\pi/3$-sectors with four domain walls emanating from the fracton core. This excitation has a lower energy of $E_{\text{1-frac}}^{\text{Potts}}=1.5J_2$ [$1.5J_{5\bigtriangleup}$, $1.5(J_{5\bigtriangleup}+J_{5\bigtriangledown})$], for the $J_1$-$J_2$-$J_{3\text{d}}$ model [$J_1$-$J_{5\bigtriangleup}$-model, $J_1$-$J_{5\bigtriangleup}$-$J_{5\bigtriangledown}$-model]. Except for the $J_1$-$J_{5\bigtriangleup}$-$J_{5\bigtriangledown}$-model, these fractons can also be moved into neighboring $J_1$ triangles via a simple shift of domain walls, illustrated in Figs.~\ref{fig:potts_fractons}(c) and \ref{fig:potts_fractons}(d). The associated excitation energies are then $4.5J_1$ and $1.5J_1$ for triple and single fractons, respectively. As a further difference between the three variants of our system the $J_1$-$J_2$-$J_{3\text{d}}$ model allows for fractons in both the upward and downward pointing $J_1$-triangles while the $J_1$-$J_{5\bigtriangleup}$ model can only host isolated fractons in downward pointing $J_1$-triangles. Up to real-space rotations around the fracton center and permutations of domains these configurations cover all isolated fractons the system may host.

Like in the plaquette Ising model, the immobility of defects is clearly established by the fact that moving a fracton requires shifting domain walls, which amounts to changing the spin configuration in the whole area swept over by the domain wall. However, as a striking difference compared to the plaquette Ising model, more than two domain boundaries are sticking out of the fracton cores. Consequently, in contrast to conventional type-I fractons models, it is impossible to create a group of isolated fractons out of a local multipole of defects such that they reside at the corners of a large region of flipped spins. In other words, exciting one or more fractons out of a ground state inevitably requires introducing domain walls reaching out to infinity.

Swapping two colors in a ${\mathbf q}=0$ state along a {\it semi-infinite} line as shown in Fig.~\ref{fig:potts_fractons}(e) creates a fracton bound state with subdimensional mobility, similar to a lineon. These excitations move by extending/shortening the string of swapped colors which amounts to shifting the blue domain in Fig.~\ref{fig:potts_fractons}(e) to the left/right. In contrast to conventional type-I fracton models this motion, however, is not strictly linear but occurs in an unusual zig-zag manner, as indicated by the light magenta stars in Fig.~\ref{fig:potts_fractons}(e). As a further difference compared to more usual fracton scenarios, the two defects forming the lineon are not of the type of isolated excitations shown in Fig.~\ref{fig:potts_fractons}(a)-(d) but rather should be considered as parts of a defective domain wall which violates the adjacency rules of Fig.~\ref{fig:potts_ground_states}(b). This becomes obvious when trying to extend the vertical thickness of the domain ``5'' in Fig.~\ref{fig:potts_fractons}(e). Indeed, there is no possible termination of this domain at its left end that obeys the adjacency rules. Hence, in contrast to the plaquette Ising model where the two defects in Fig.~\ref{fig:plaquette_ising}(c) may be vertically separated without energy cost, such a separation would lead to a string of defects in Fig.~\ref{fig:potts_fractons}(e) with an energy proportional to its length. Reversely, a subdimensional excitation consisting of two isolated defects, each of the type of Fig.~\ref{fig:potts_fractons}(a)-(d) does not exist.

\subsubsection{Interpretation of the fracton behaviors}
As discussed above our kagome Potts models show unusual fracton behaviors: $(i)$ Fractons cannot be created at the corners of a finite region of modified spins and (ii) two-fracton bound states move along zig-zag paths and cannot be extended in a direction parallel to themselves. This is in contrast to conventional type-I fracton models~\cite{vijay16,jack05,lipowski97,garrahan02} where fractons are created as a group of four at the corners of a rectangular region. Such differences present an interesting challenge of how to interpret them in a natural and unifying manner.

The following considerations apply to the $J_1$-$J_2$-$J_{3\text{d}}$ model but their implications also hold for the other two models. Furthermore, the arguments below do not depend on the existence of three colors. For simplicity, it suffices to consider the degrees of freedom arising from interchanging, e.g., red and blue colors. Starting from a ground state and changing a single red into a blue spin means that the two adjacent triangles and hexagons obey $\sum_{i\in\text{triangle}}{\mathbf S}_i=\sum_{i\in\text{hexagon}}{\mathbf S}_i={\mathbf S}_\text{blue}-{\mathbf S}_\text{red}\equiv +{\mathbf d}$. The reverse process leads to two triangles and two hexagons with spin sums $-{\mathbf d}$ such that $\pm{\mathbf d}$ are effective $\mathbb{Z}_2$ charges.

\begin{figure}
	\includegraphics[width=0.9\columnwidth]{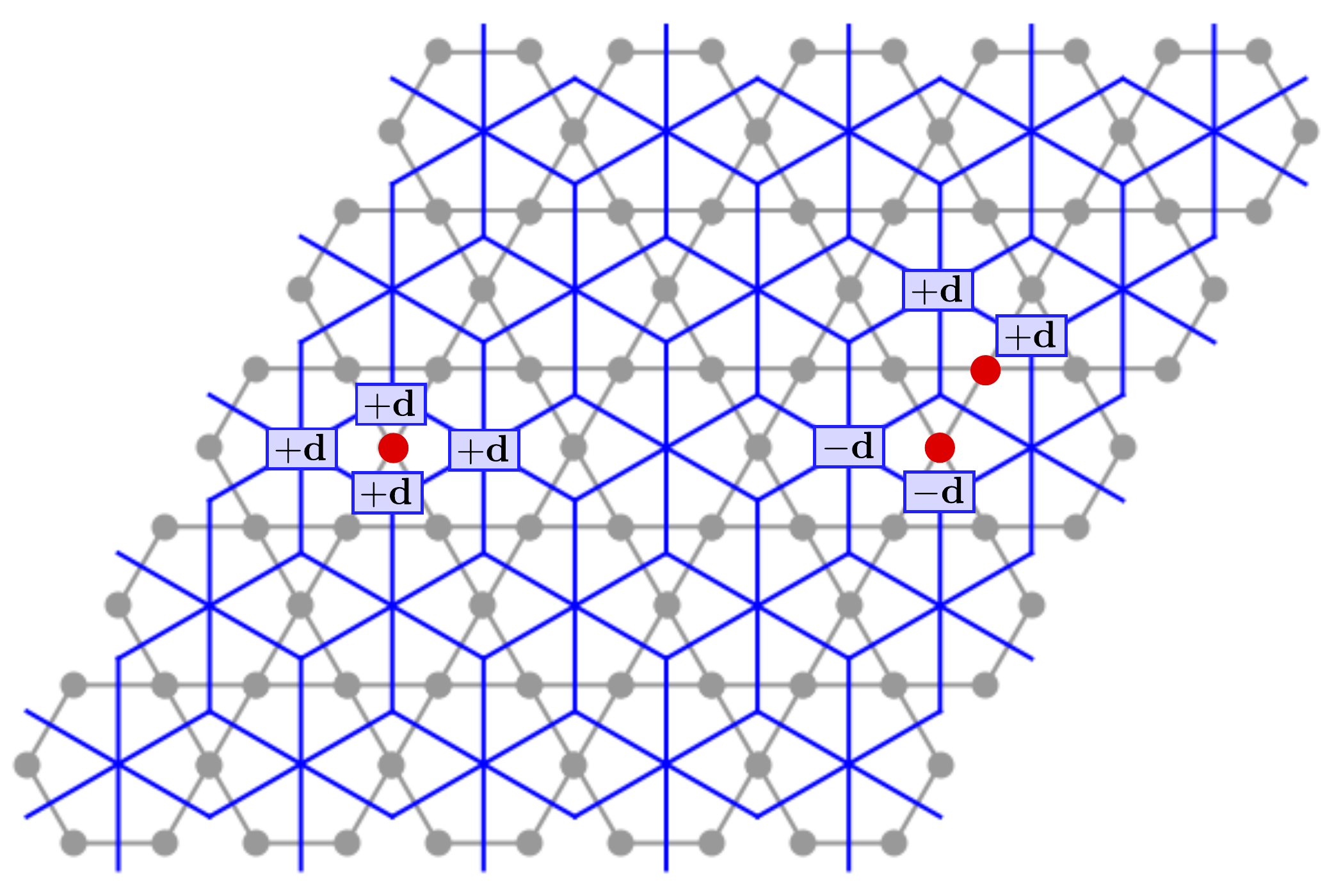}
	\caption{The rhombille tiling (blue lattice) and kagome lattice (gray lattice) are dual to each other. Left: Quadrupole resulting from a single spin flip. Right: Charge configuration from two neighboring $+{\mathbf d}$ and $-{\mathbf d}$ quadrupoles. Modified spins are indicated by red points.\label{fig:Rhombille}}
\end{figure}

Conforming to the rank-2 U(1) gauge theory interpretation of fractons~\cite{pretko17,pretko17_2}, we can think of the spin degrees of freedom sitting on the sites of the kagome lattice  as  discrete  electric fields. The color-neutrality conditions on the triangles and hexagons corresponds to a ``Gauss's law'' which relates the electric field to the ``electric charges'' (i.e., defect triangles/hexagons). Hence, the fractons reside on the dual lattice of the kagome lattice which is formed by the centers of the triangles and hexagons, called the rhombille tiling (cf. Fig.~\ref{fig:Rhombille}). Reversely, the spin degrees of freedom, or electric fields, live on the centers of the rhombi. Changing the value of one spin creates a quadrupole of four excitations with charges $\pm{\mathbf d}$ on the corners of a rhombus, see Fig.~\ref{fig:Rhombille} (left). The fracton behaviors now crucially depend on how the lattice can be tiled with these quadrupoles. As an example we show in Fig.~\ref{fig:Rhombille} (right) a charge configuration resulting from a pair of neighboring $+{\mathbf d}$ and $-{\mathbf d}$ quadrupoles.

A first indication for the unusual fracton behavior comes from the fact that the rhombille tiling is lacking ``scale invariance''. Particularly, enlarged versions of the elementary rhombi such as parallelograms extending over large regions do not exist in the lattice. This is in contrast to fracton models on square or cubic lattices where rectangular areas of any size can be embedded into the lattice. Below, these ideas will be formulated in a more rigorous way.

We can view the rhombille tiling as a two-layer cut of a cubic lattice, perpendicular to the $[1,1,1]$ direction. In this representation, each rhombus is a face of an elementary cube. Hence, the fracton quadrupole created by changing the value of a single spin now lives on the corners of a square embedded in a square lattice in either the $x$-$y$, $y$-$z$, or $x$-$z$ plane. This is the familiar type-I fracton scenario as, e.g., realized in the plaquette Ising model~\cite{jack05,lipowski97,garrahan02}. By considering the full cubic lattice, large rectangular regions in the $x$-$y$, $y$-$z$, or $x$-$z$ planes can be tiled with quadrupoles which creates the usual type of isolated fractons. Furthermore, a dipole of fractons can move in the plane perpendicular to itself.
\begin{figure}
	\centering
	\subfloat[\label{fig:Rhombille.fracton.1}]{\includegraphics[width=0.45\columnwidth]{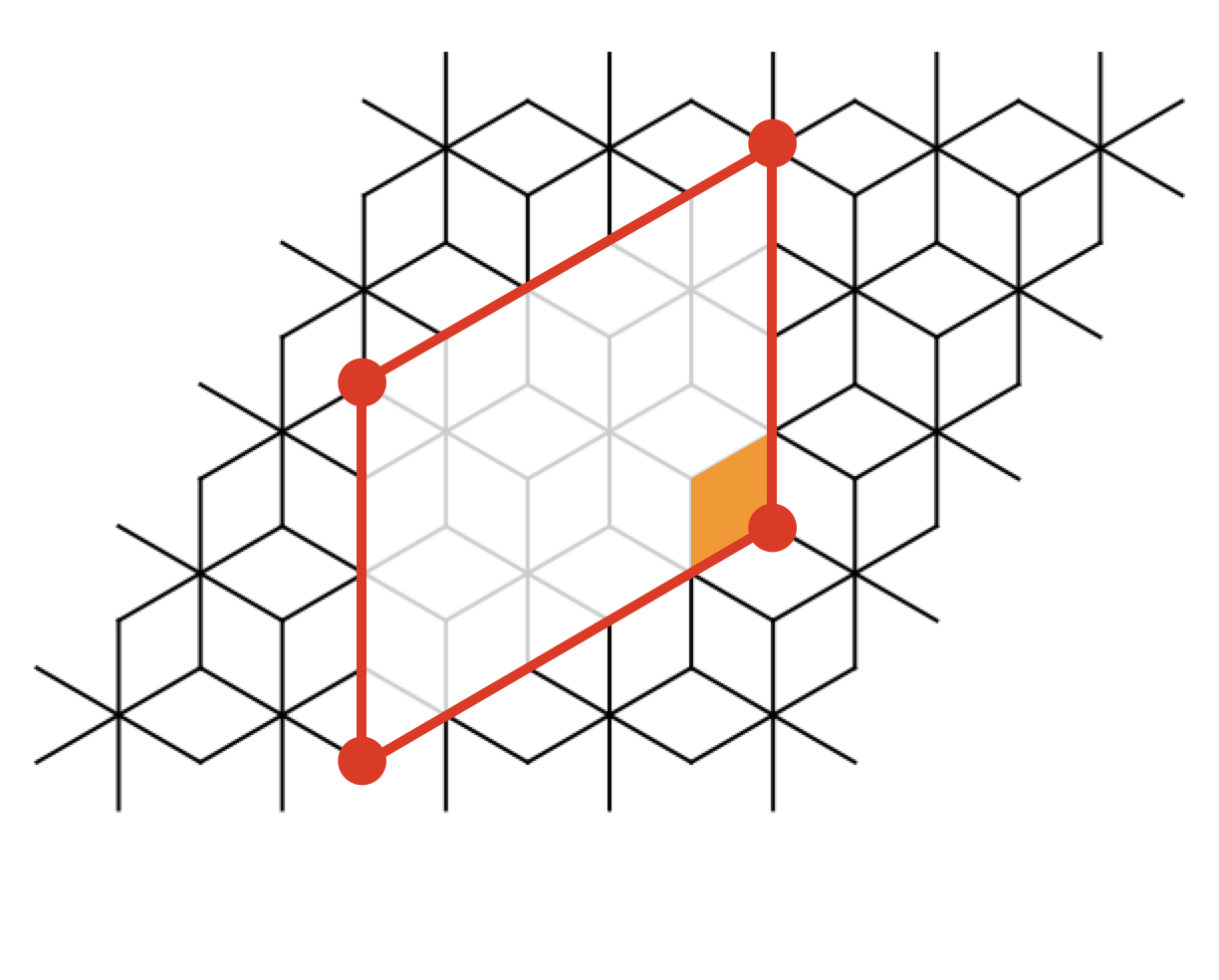}}\;
	\subfloat[\label{fig:Rhombille.fracton.2}]{\includegraphics[width=0.45\columnwidth]{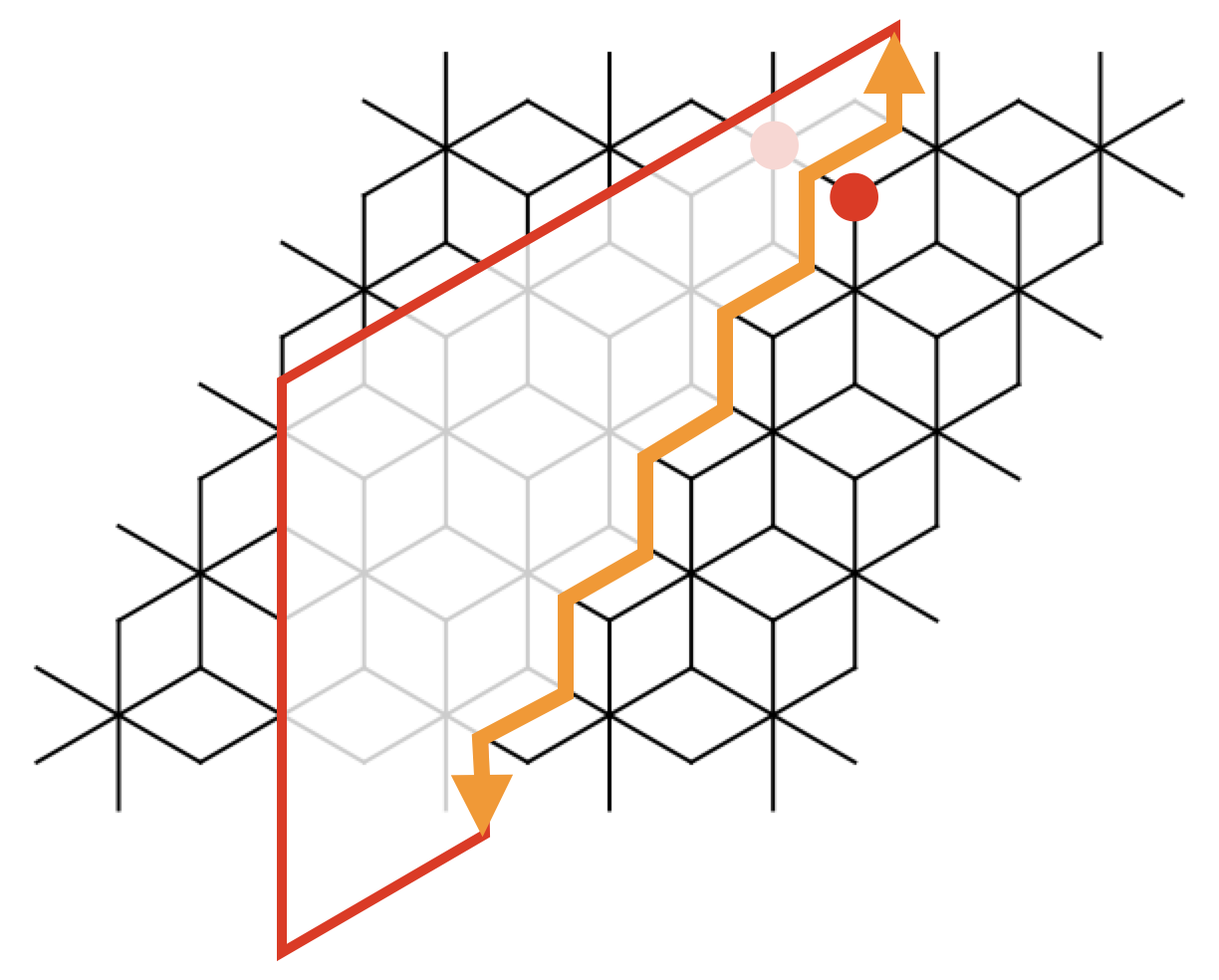}}\;
	\caption{
		Fracton behaviors explained by viewing the rhombille tiling embeded in the cubic lattice.
		(a) A large rectangle highlighted in red with a quadrupole of fractons on its corners does not fully overlap with the rhombille tiling cut. 
		Hence a single fracton originating from a far-separated quadrupole cannot be created on the rhombille tiling.
		(b) A fracton dipole can move in the plane perpendicular to itself in the cubic latice. 
		This plane only intersects the rhombille tiling cut on the one-dimensional orange path. 
		Hence a fracton dipole on the rhombille tiling is restricted to move along the orange zig-zag path.
	}
	\label{fig:Rhombille.fracton}
\end{figure}

The fracton behavior on the rhombille tiling can, hence, be explained by familiar fracton type-I behavior on a cubic lattice, but restricted to the two-layer cut forming the rhombille tiling. This means that any quadrupoles outside the two-layer cut are not usable. As a consequence, a large rectangular region in the $x$-$y$, $y$-$z$, or $x$-$z$ plane does not live entirely on the cut in $[1,1,1]$ direction and, hence, one cannot create isolated fractons from a far-separated quadrupole, see Fig.~\ref{fig:Rhombille.fracton.1}. Additionally, as illustrated in Fig.~\ref{fig:Rhombille.fracton.2}, a plane perpendicular to a fracton dipole intersects the $[1,1,1]$ cut on a one-dimensional sub-manifold, forming a zig-zag path on the rhombille tiling in agreement with our observation in Fig.~\ref{fig:potts_fractons}(e). This dipole cannot be extended beyond a separation of one lattice spacing since this would lead to fractons outside the $[1,1,1]$ cut.

\subsubsection{Low energy effective theory}
In the next step, we discuss the low energy effective theory underlying  the observed fracton behaviors. The following considerations are based on a simple cubic lattice in which the kagome model can be embedded. The cubic lattice hosts a rank-2 U(1) electrostatics theory~\cite{pretko17,pretko17_2}, where the rank-2 electric field is a symmetric tensor with all diagonal components vanishing:
\begin{equation}
    {\bf E} = 
    \begin{bmatrix}
    0 & E^{xy} & E^{zx} \\
    E^{xy} & 0 & E^{yz} \\
    E^{zx} & E^{yz} & 0 \\
    \end{bmatrix}   .
\end{equation}
This type of rank-2 U(1) theory
is found to be
crucial in deriving three-dimensional gapped
fracton topological order
by Higgsing gapless rank-2 U(1) theories \cite{bulmash18_2,ma18}.
Note that the components $E^{ij}$ are defined on different lattice positions. For example, denoting the cubic lattice sites by ${\mathbf r}$, the component $E^{xy}$ lives on the centers of elementary plaquettes in the $x$-$y$ plane, i.e., at ${\mathbf r}+(\pm{\mathbf e}_x\pm{\mathbf e}_y)/2$ (where ${\mathbf e}_i$ are cartesian unit vectors). The other components are defined equivalently. The Gauss's law describing the charge-free sector is that of a scalar charged rank-2 U(1) theory:
\begin{equation}
    \partial_i \partial_j E^{ij} = 0    .\label{gauss}
\end{equation}
To connect this condition to our above discussion, we formulate a discretized lattice version of Eq.~(\ref{gauss}) which exists for all cubic sites ${\mathbf r}$,
\begin{equation}
E^{ij}_{{\mathbf r}+\frac{{\mathbf e}_i}{2}+\frac{{\mathbf e}_j}{2}}-E^{ij}_{{\mathbf r}+\frac{{\mathbf e}_i}{2}-\frac{{\mathbf e}_j}{2}}-E^{ij}_{{\mathbf r}-\frac{{\mathbf e}_i}{2}+\frac{{\mathbf e}_j}{2}}+E^{ij}_{{\mathbf r}-\frac{{\mathbf e}_i}{2}-\frac{{\mathbf e}_j}{2}}=0\;.
\end{equation}
Hence, changing $E^{ij}$ on a single plaquette in a state that satisfies all constraints creates a quadrupole of defects at the corners of the plaquette. These quadrupoles correspond to the ones illustrated in Fig.~\ref{fig:Rhombille} for the original kagome lattice. The fact that there are only off-diagonal components in the field tensor is in agreement with the observation that only quadrupoles on a plaquette can be created by changing $E^{ij}$. The low-energy effective Hamiltonian can, therefore, be written as
\begin{equation}
    H=U(\partial_i \partial_j E^{ij})^2\;,
\end{equation}
penalizing the existence of a charge with an energy cost. Since there are only off-diagonal components of $E^{ij}$
living on the faces of the cubes, 
the simplest way to vary $E^{ij}$ while respecting the Gauss' laws
is to shift all $E^{ij}$ by the same value on an entire plane.
Degeneracies of line-  or local spin flips do not exist. In the picture of multipoles, this amounts to extending the  quadrupole on a plaquette to infinity on the plane it lives on.

The low-energy physics of the kagome model can again be interpreted as the restriction of the above 3D tensor gauge theory to a $[1,1,1]$ cut. Particularly, the degeneracies from flipping three types of planes in the 3D system are translated into three types of line degeneracies on the kagome lattice.
The fracton excitations can also be consistently explained in the same fashion, as we discussed in detail in the previous section.
\begin{figure}
\includegraphics[width=0.9\linewidth]{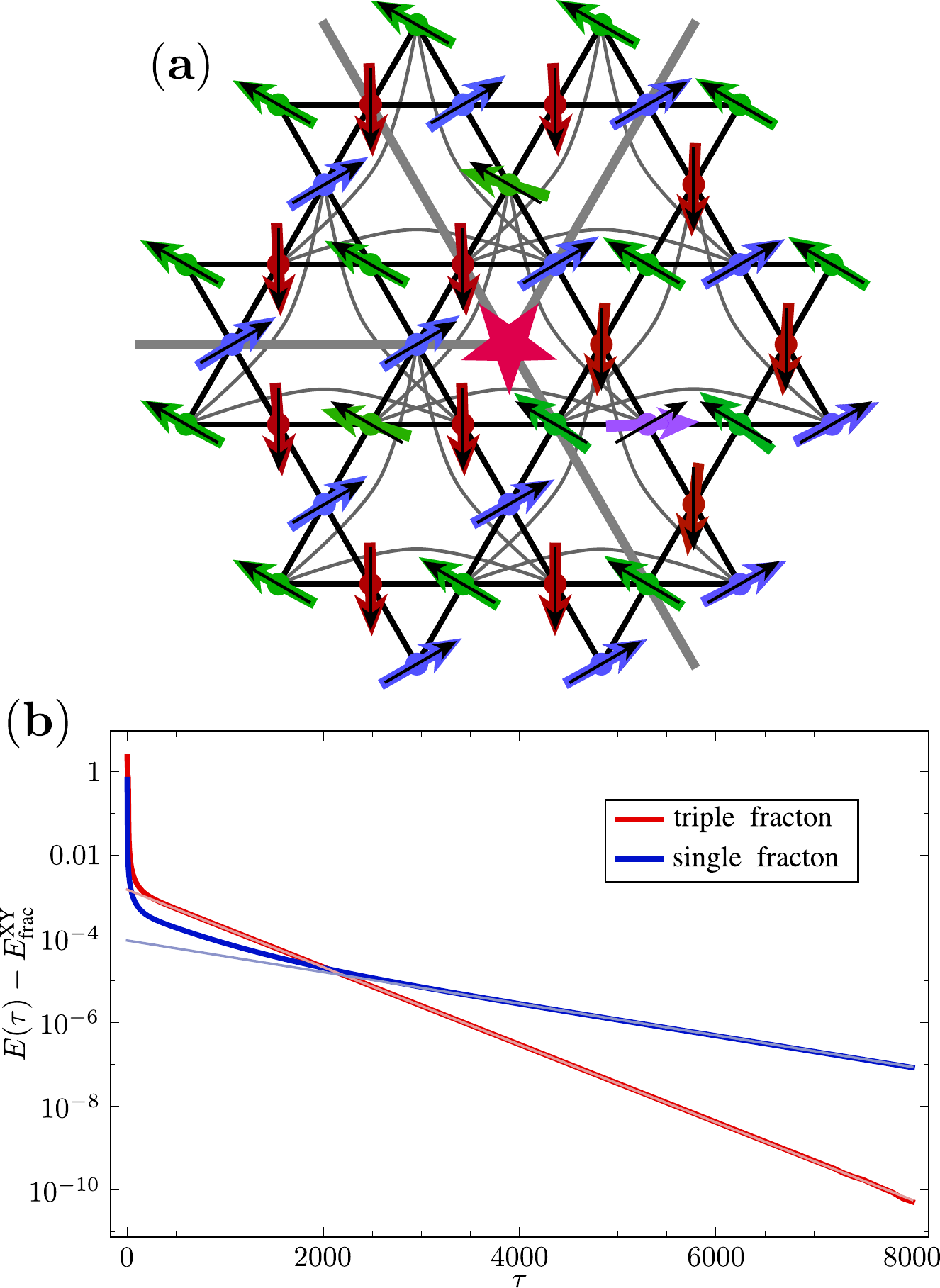}
\caption{(a) Optimized spin configuration of a single fracton (magenta star) in the $J_1$-$J_{5\bigtriangleup}$ model from iterative minimization. Small black arrows at each lattice point illustrate the initial defect of the three-state Potts model. (b) Relaxation of excitation energies $E(\tau)-E_{\text{1-frac}}^{\text{XY}}$ and $E(\tau)-E_{\text{3-frac}}^{\text{XY}}$ as a function of the step count $\tau$ for a single and triple defect. Thin lines are fits to an exponential function.\label{fig:xy_fracton_optimze}}
\end{figure}

\subsection{XY models}\label{xy_low_energy}
We now generalize our system from three-state Potts spins to continuous in-plane XY spins. Particularly, we discuss the associated modifications of ground states and isolated defect states from the last section, which turn out to be rather small (new phenomena, however, emerge when considering vortex states, see Sec.~\ref{numerics_xy}). Apart from the freedom to globally rotate spins within the $x$-$y$ plane, the ground states are the same as in the three-state Potts model. Likewise, there exist two types of fractons for all three models ($J_1$-$J_2$-$J_{3\text{d}}$, $J_1$-$J_{5\bigtriangleup}$, and $J_1$-$J_{5\bigtriangleup}$-$J_{5\bigtriangledown}$ models) which we again call single and triple fractons. An obvious difference, however, is that the optimal spin configurations in the cores of defects (which correspond to a local energy minimum) are slightly deformed compared to the fractons in the Potts model. Such optimized states can be most easily constructed by performing an iterative minimization scheme~\cite{walker80,sklan13}: The starting configuration is a defect from the three-state Potts model in the center of a system with open boundary conditions. We then successively select random spins and orient them along its so-called local field ${\mathbf h}_i$ which, for a general Hamiltonian $H=\sum_{\langle i,j\rangle}J_{ij}{\mathbf S}_i\cdot{\mathbf S}_j$ is defined by
\begin{equation}
{\mathbf h}_i=-\sum_j J_{ij}{\mathbf S}_j\;.
\end{equation}   
By construction, in each such step the energy can only be lowered, however, as a steepest descent method the scheme can get stuck in a local minimum (which in our case is a fracton state). An example of an optimized single fracton in the $J_1$-$J_{5\bigtriangleup}$ model with $J_1=J_{5\bigtriangleup}$ is illustrated in Fig.~\ref{fig:xy_fracton_optimze}(a), together with the initial spin configuration of a fracton in the three-state Potts model (small black arrows). While both spin configurations in Fig.~\ref{fig:xy_fracton_optimze}(a) closely resemble each other, the excitation energies are reduced quite significantly, from $E_{\text{1-frac}}^{\text{Potts}}=1.5J_1$ to $E_{\text{1-frac}}^{\text{XY}}=0.8546J_1$ for a single fracton and from $E_{\text{3-frac}}^{\text{Potts}}=4.5J_1$ to $E_{\text{3-frac}}^{\text{XY}}=2.2014J_1$ for a triple fracton. Furthermore, the real space distribution of excitation energies is no longer given by a single defect triangle but spreads over a few lattice spacings. To estimate the spatial extent of the fracton in Fig.~\ref{fig:xy_fracton_optimze}(a) we note that at a distance of $4$ lattice spacings away from its center, the spins do not deviate more than $4.6^\circ$ compared to the initial configuration. Adopting the concept of fugacity as a measure for the size of vortex cores in Kosterlitz-Thouless systems, one may conclude that fractons in the three-state Potts model have the smallest possible fugacity which increases for the XY model.

We also note that the decrease of the excitation energy as a function of the step count $\tau$ in iterative minimization follows a standard exponential relaxation at sufficiently large $\tau$ (here one step corresponds to $N=7500$ individual updates, where $N$ is the total number of lattice sites). This is shown in  Fig.~\ref{fig:xy_fracton_optimze}(b), where the decay of the excitation energies $E(\tau)-E_{\text{1-frac}}^{\text{XY}}$ and $E(\tau)-E_{\text{3-frac}}^{\text{XY}}$ is plotted as a function of $\tau$. As will be discussed in the next section, this behavior is in stark contrast to the power-law relaxation of defects in the Heisenberg models.

\subsection{Heisenberg models}
We now consider the ground states and low-energy defect states of the Heisenberg models. Since the $J_1$-$J_2$-$J_{3\text{d}}$, $J_1$-$J_{5\bigtriangleup}$, and $J_1$-$J_{5\bigtriangleup}$-$J_{5\bigtriangledown}$ models all have very similar properties (the degenerate ground states are even identical in the three cases) they can be treated together.

\subsubsection{Ground states of the Heisenberg models}\label{sec:ground_heisenberg}
It is clear that all ground states of the three color models are also ground states of the Heisenberg models, however, the reverse is not true, i.e. not all ground states of the Heisenberg models consist of only three spin orientations. Indeed, as demonstrated below, the set of degenerate ground states of the Heisenberg models is surprisingly rich and interesting on its own. Apart from global $SO(3)$ rotations of all spins, the ground state manifold can be constructed by subsystem operations. This is similar to the three color models where, starting from a ${\mathbf q}=0$ state, one can generate all ground states by swapping two colors on arbitrary parallel lines. In contrast to the three color models, however, the Heisenberg models also allow for subsystem operations on lines {\it which cross}. The lines on which such operations are performed can have three orientations and one may classify each ground state according to the number of different directions of line manipulations required to generate it out of a ${\mathbf q}=0$ state. The first case $(i)$ is similar to the three color models, i.e., only operations along lines with the same lattice direction are involved. In the second case $(ii)$, manipulations along two different lattice directions are required and in the third case $(iii)$, all three types of operations are performed. Below, we will consider the ground states generated in each of the three cases and show that they are characterized by different types of discrete and/or continuous degeneracies.
\begin{figure}
\includegraphics[width=0.5\linewidth]{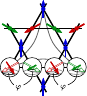}
\caption{Illustration of line-manipulations in the ground states of case $(i)$: All spins along a line with alternating green and red colors are rotated by an angle $\varphi$ around the direction of the blue spins. Arrows indicated by contours show the orientations of the spins before the transformation.\label{fig:heis_ground}}
\end{figure}

{\it Case $(i)$: Subsystem operations on parallel lines.} This case is a direct generalization of the construction of ground states in the three color models. The operation of swapping two colors along a line in the latter models, which generates all ground states, corresponds to performing a $\pi$-rotation around the direction of the third spin. In a model with $SO(3)$ spins, this rotation can be performed by any angle $\varphi\in[0,2\pi)$ without exciting the system, as illustrated in Fig.~\ref{fig:heis_ground}. Since one can independently do such rotations on arbitrary numbers of parallel lines where each subsystem operation is characterized by an angle $\varphi$ the ground state manifold is defined by $L$ continuous parameters. In all states generated this way, one of the kagome sublattices has a fixed direction.

\begin{figure}
\includegraphics[width=0.9\linewidth]{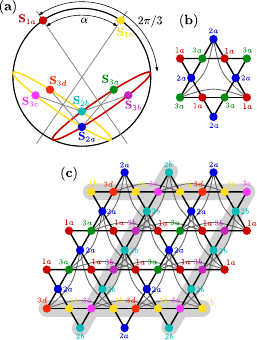}
\caption{Construction of degenerate ground states in case $(ii)$: (a) The dots denote the Bloch sphere locations of the eight spins defining these states. The angle $\alpha\in[0,2\pi/3)$ is a free parameter. The red (yellow) circle indicates all positions on the Bloch sphere enclosing an angle of $2\pi/3$ with the red (yellow) spin. Spins with opposite positions on the red and yellow circles are connected by gray lines. The ${\mathbf q}=0$ state in (b) represents the starting configuration for the manipulations performed in the example in (c). The lines on which the mirror operations described in the text act are highlighted by a gray shaded background in (c).\label{fig:two_lines}}
\end{figure}

{\it Case $(ii)$: Subsystem operations on lines with two directions.} Apart from global $SO(3)$ rotations, the degenerate ground states in this case depend on one continuous angle $\alpha\in[0,2\pi/3)$ and are constructed using eight different spin orientations which we denote ${\mathbf S}_{1a}$, ${\mathbf S}_{1b}$, ${\mathbf S}_{2a}$, ${\mathbf S}_{2b}$, ${\mathbf S}_{3a}$ ${\mathbf S}_{3b}$, ${\mathbf S}_{3c}$, ${\mathbf S}_{3d}$. The free parameter $\alpha$ is the angle between the spins ${\mathbf S}_{1a}$ and ${\mathbf S}_{1b}$ which we assume to be fixed in the following (changing ${\mathbf S}_{1a}$ and ${\mathbf S}_{1b}$ amounts to changing the angle $\alpha$ and/or performing a global $SO(3)$ rotation of all spins). The other spin orientations are then defined according to Fig.~\ref{fig:two_lines}(a): Drawing a circle on the Bloch sphere around ${\mathbf S}_{1a}$ such that all points on the circle enclose an angle of $2\pi/3$ with ${\mathbf S}_{1a}$ and drawing the same type of circle around ${\mathbf S}_{1b}$, the spins ${\mathbf S}_{2a}$ and ${\mathbf S}_{2b}$ point in the two directions where the circles cross. The remaining spin directions ${\mathbf S}_{3a}$ ${\mathbf S}_{3b}$, ${\mathbf S}_{3c}$, ${\mathbf S}_{3d}$ lie on opposite positions on the two circles, as specified in Fig.~\ref{fig:two_lines}(a). For example, ${\mathbf S}_{3a}$ lies on the circle around ${\mathbf S}_{1a}$ opposite to ${\mathbf S}_{2a}$ such that ${\mathbf S}_{2a}$ and ${\mathbf S}_{3a}$ enclose an angle of $2\pi/3$.

We construct the degenerate ground states in this case by starting from a ${\mathbf q}=0$ state. For the three spin orientations defining the ${\mathbf q}=0$ state we select an arbitrary trio of spins from the eight directions, such that, in the usual way, each pair of the trio encloses an angle of $2\pi/3$. In the example considered here [see Fig.~\ref{fig:two_lines}(b)], we choose the three spins ${\mathbf S}_{1a}$, ${\mathbf S}_{2a}$, and ${\mathbf S}_{3a}$ (it is clear from their construction that all three numbers $``1"$, $``2"$, $``3"$ must be represented once in the index labels of such a trio). Manipulations of the state in Fig.~\ref{fig:two_lines}(b) can now be independently performed along the two types of lines where spins ${\mathbf S}_{1\bullet}$, ${\mathbf S}_{3\bullet}$, ${\mathbf S}_{1\bullet}$, $\ldots$ alternate or where ${\mathbf S}_{2\bullet}$, ${\mathbf S}_{3\bullet}$, ${\mathbf S}_{2\bullet}$, $\ldots$ alternate (here `$\bullet$' is a placeholder for $a$, $b$, $c$, $d$). Particularly, an operation on a line with ${\mathbf S}_{1\bullet}$, ${\mathbf S}_{3\bullet}$, ${\mathbf S}_{1\bullet}$, $\ldots$ (${\mathbf S}_{2\bullet}$, ${\mathbf S}_{3\bullet}$, ${\mathbf S}_{2\bullet}$, $\ldots$) amounts to mirror all spins on the Bloch sphere with respect to the plane passing through ${\mathbf S}_{2a}$, ${\mathbf S}_{2b}$ and the origin (passing through ${\mathbf S}_{1a}$, ${\mathbf S}_{1b}$ and the origin). Since these mirror operations commute, the order in which the line-manipulations are performed is irrelevant. Again, this specific construction ensures that no such operation excites the system. An example of a spin configuration resulting from manipulating the ${\mathbf q}=0$ state in Fig.~\ref{fig:two_lines}(b) is shown in Fig.~\ref{fig:two_lines}(c) where modified lines are indicated by a gray shaded background.

Like in case $(i)$ the mirror operations can be described as $\varphi$-rotations around the direction of the third spin. Particularly, consider a manipulated nearest neighbor pair of spins on a line  ${\mathbf S}_{1\bullet}$, ${\mathbf S}_{3\bullet}$, ${\mathbf S}_{1\bullet}$, $\ldots$ (${\mathbf S}_{2\bullet}$, ${\mathbf S}_{3\bullet}$, ${\mathbf S}_{2\bullet}$, $\ldots$). This pair is connected via $J_1$ to a spin ${\mathbf S}_{2\bullet}$ (${\mathbf S}_{1\bullet}$) outside the chain. The manipulation of this pair is then a result of rotating it around ${\mathbf S}_{2\bullet}$ (${\mathbf S}_{1\bullet}$) by an angle $\varphi_2$ ($\varphi_1$) given by
\begin{align}
&\varphi_1=\arccos\left(\frac{-1-5\cos\alpha}{3+3\cos\alpha}\right)\;,\notag\\
&\varphi_2=\arccos\left(-\frac{1}{3}+\frac{4}{3}\cos\alpha\right)\;.
\end{align}
In total, since the described operations can be independently performed along each of the two types of lines, up to a global rotation of all spins the degeneracy in this case results from choosing a continuous angle $\alpha$ {\it and} performing $(2^L)^2$ possible discrete operations. 
\begin{figure}
\includegraphics[width=0.9\linewidth]{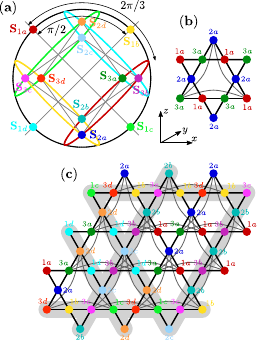}
\caption{Construction of degenerate ground states in case $(iii)$: (a) Location of the twelve spin directions $(\pm1,\pm1,0)/\sqrt{2}$, $(\pm1,0,\pm1)/\sqrt{2}$, $(0,\pm1,\pm1)/\sqrt{2}$ on the Bloch sphere where we fixed the global orientation. The colored circles enclose an angle of $2\pi/3$ with the spin of the same color. Spins with opposite positions on these circles are connected by gray lines. The ${\mathbf q}=0$ state in (b) represents the starting configuration for the manipulations performed in the example in (c). The lines on which the mirror operations described in the text act are highlighted by a gray shaded background in (c).\label{fig:three_lines}}
\end{figure}

{\it Case $(iii)$: Subsystem operations on lines with all three directions.} The construction of degenerate ground states in this case is similar to case $(ii)$, however, the angle $\alpha$ is fixed to $\alpha=\pi/2$ such that (apart from global rotations) no continuous free parameter exists. In addition to the eight spin orientations ${\mathbf S}_{1a}$, ${\mathbf S}_{1b}$, ${\mathbf S}_{2a}$, ${\mathbf S}_{2b}$, ${\mathbf S}_{3a}$ ${\mathbf S}_{3b}$, ${\mathbf S}_{3c}$, ${\mathbf S}_{3d}$ from case $(ii)$ the system may host four more spins given by ${\mathbf S}_{1c}=-{\mathbf S}_{1a}$, ${\mathbf S}_{1d}=-{\mathbf S}_{1b}$, ${\mathbf S}_{2c}=-{\mathbf S}_{2a}$, ${\mathbf S}_{2b}=-{\mathbf S}_{2d}$, see Fig.~\ref{fig:three_lines}(a). Fixing the global orientation, we may choose the ${\mathbf S}_{1\bullet}$-spins to lie in the $x$-$z$ plane at positions $(\pm1,0,\pm1)/\sqrt{2}$. Equivalently, the ${\mathbf S}_{2\bullet}$-spins (the ${\mathbf S}_{3\bullet}$-spins) reside in the $y$-$z$ plane ($x$-$y$ plane) at $(0,\pm1,\pm1)/\sqrt{2}$ ($(\pm1,\pm1,0)/\sqrt{2}$). We again start the construction of degenerate spin configurations with the ${\mathbf q}=0$ state shown in Fig.~\ref{fig:three_lines}(b) which is based on ${\mathbf S}_{1a}$, ${\mathbf S}_{2a}$, and ${\mathbf S}_{3a}$-spins (note that any other choice of a planar trio of spins with mutual angles $2\pi/3$ can, likewise, be used for the initial configuration). Line-manipulations can now be independently performed along all three lattice directions: An operation on a line ${\mathbf S}_{1\bullet}$, ${\mathbf S}_{2\bullet}$, ${\mathbf S}_{1\bullet}$, $\ldots$ amounts to mirror the spins with respect to the plane in which the ${\mathbf S}_{3\bullet}$-spins reside (i.e. the $x$-$y$ plane). The definitions of the other two types of line-manipulations follow by cyclic permutations of indices $1\rightarrow2\rightarrow3\rightarrow1$. In Fig.~\ref{fig:three_lines}(c) we show an example of a spin configuration obtained after various such operations. We may, alternatively, view the transformation of any pair of nearest neighbor spins on a modified line as a rotation by $\varphi=\arccos(-1/3)\approx109.5^\circ$ around the  direction of the third spin in the respective $J_1$-triangle. Again, all manipulations are independent of each other (i.e., they commute) such that the total discrete degeneracy in case $(iii)$ is proportional to $(2^L)^3$.

In summary, the degenerate ground states in our kagome Heisenberg models fall into three classes which are characterized by the type of allowed subsystem operations within each class. In the simplest case $(i)$ continuous spin rotations on any set of parallel lines can be performed. In contrast, case $(ii)$ and $(iii)$ also permit subsystem operations on intersecting lines which, however, comes at the expense that such manipulations are only of discrete nature. As we will see below, these ground state degeneracies also have consequences for the properties of excited defect states.
\begin{figure}
\includegraphics[width=0.99\linewidth]{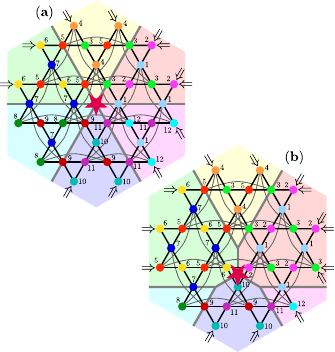}
\caption{Construction of isolated defects (illustrated by magenta stars) located (a) in the center of a hexagon and (b) in the center of a $J_1$-triangle. The configurations consist of six different ${\mathbf q}=0$ domains (indicated by light background colors) which involves twelve different spin orientations (indicated by numbers $1-12$ and by different colors of the points). The arrows $\Rightarrow\cdots\Leftarrow$ mark the lines which need to be modified according to case $(iii)$ in Sec.~\ref{sec:ground_heisenberg} to generate a zero energy defect out of a ${\mathbf q}=0$ state.\label{fig:heis_fracton}}
\end{figure}

\subsubsection{Single defects in the Heisenberg models}
Knowing the ground states of the Heisenberg models we continue discussing the possible existence of fracton-like excited states. The procedure here is similar to the three-state Potts and XY models, i.e., we first construct `by hand' states with a single isolated defect triangle or hexagon (which generalizes the single and triple fractons discussed above). We then investigate their stability by applying iterative minimization. As we will see below, fracton-like excitations have nearly identical properties for all three models ($J_1$-$J_2$-$J_{3\text{d}}$, $J_1$-$J_{5\bigtriangleup}$, and $J_1$-$J_{5\bigtriangleup}$-$J_{5\bigtriangledown}$ models) such that they are again treated together.

The simplest and most generic way of constructing a single defect (which applies to all three systems) is by merging six sectors with different ${\mathbf q}=0$ states as illustrated by the colored regions in Fig.~\ref{fig:heis_fracton}(a). The defect is then located in the center hexagon where the domain walls separating the six sectors intersect. All other hexagons and small/large triangles not part of this center hexagon shall fulfill the local constraints. Since neighboring sectors share one type of spin, these configurations have a maximum twelve different spin orientations labelled $1-12$ in Fig.~\ref{fig:heis_fracton}(a).

A priori it is not clear whether solutions other than the single and triple fractons from the three state Potts model exist. However, simple geometric considerations show that up to a global $SO(3)$ rotation of all spins a continuous manifold of defects exists which may be characterized by three continuous parameters  $u$, $v$, $w$, each in the interval $u,v,w\in[-1,1]$. One possible way of parametrizing the twelve (normalized) spins ${\mathbf S}_1,\ldots,{\mathbf S}_{12}$ in terms of $u$, $v$, $w$ (where $u=0$ must be excluded) is given by
\begin{equation}
{\mathbf S}_1=\left(\begin{array}{c}0\\0\\1\end{array}\right)\;,\;{\mathbf S}_5=\left(\begin{array}{c}\sin(\frac{2\pi |u|}{3})\\0\\\cos(\frac{2\pi u}{3})\end{array}\right)\;,\;{\mathbf S}_9=c\,{\mathbf e}+{\mathbf d}\;,
\end{equation}
where
\begin{align}
&{\mathbf d}=-\frac{1}{2}\left(\begin{array}{c}\tan(\frac{\pi|u|}{3})\\0\\1\end{array}\right)\;,\notag\\
&{\mathbf e}=\left(\begin{array}{c}\sin(\pi|v|)\sin\left[\frac{2\pi|u|}{3}-\frac{\pi}{2}+|w|\left(\pi-\frac{2\pi|u|}{3}\right)\right]\\\cos(\pi v)\\\sin(\pi|v|)\cos\left[\frac{2\pi|u|}{3}-\frac{\pi}{2}+|w|\left(\pi-\frac{2\pi|u|}{3}\right)\right]\end{array}\right)\;,\notag\\
&c=-{\mathbf d}\cdot{\mathbf e}+\sqrt{({\mathbf d}\cdot{\mathbf e})^2-\frac{1}{4\cos^2(\frac{\pi u}{3})}+1}\;.
\end{align}
Furthermore,
\begin{align}
{\mathbf S}_i&=-\frac{{\mathbf S}_{i+2}+{\mathbf S}_{i-2}}{2(1+{\mathbf S}_{i+2}\cdot{\mathbf S}_{i-2})}+\text{sgn}(\xi_i)\,{\mathbf S}_{i+2}\times{\mathbf S}_{i-2}\,\cdot\notag\\
&\cdot\sqrt{\frac{(1+2{\mathbf S}_{i+2}\cdot{\mathbf S}_{i-2})}{2(1+{\mathbf S}_{i+2}\cdot{\mathbf S}_{i-2})\left[1-({\mathbf S}_{i+2}\cdot{\mathbf S}_{i-2})^2\right]}}\;,
\end{align}
with $i\in\{3,7,11\}$, $\xi_3=u$, $\xi_7=v$, $\xi_{11}=w$ and ${\mathbf S}_{13}\equiv{\mathbf S}_{1}$. Finally,
\begin{equation}
{\mathbf S}_j=-\frac{{\mathbf S}_{j+1}+{\mathbf S}_{j-1}}{2(1+{\mathbf S}_{j+1}\cdot{\mathbf S}_{j-1})}\;,
\end{equation}
with $j\in\{2,4,6,8,10,12\}$. We note that $u$, $v$, $w$ cover each possible isolated defect exactly once. The single and triple fractons of the three-state Potts model discussed in Sec.~\ref{sec:analytic_potts} are special cases which correspond to $\{u=0^+,v=-1/2,w=1/2\}$ and $\{u=0^+,v=-1/2,w=-1/2\}$, respectively.

Note that for the $J_1$-$J_2$-$J_{3\text{d}}$ model, the defect can also be shifted into a $J_1$ triangle (either an up-pointing or a down-pointing one) by shifting one domain wall, as illustrated in Fig.~\ref{fig:heis_fracton}(b). In the case of the $J_1$-$J_{5\bigtriangleup}$ model the single defect can only be moved into a downward pointing $J_1$ triangle [see Fig.~\ref{fig:heis_fracton}(b)] while this is not possible for an upward pointing $J_1$ triangle. Finally, the $J_1$-$J_{5\bigtriangleup}$-$J_{5\bigtriangledown}$ model cannot host any isolated defects in a $J_1$ triangle.

We also note that for isolated defects with special values of $u$, $v$, $w$ it is possible to perform additional manipulations of spins on line-like subsystems without energy cost. The construction is similar to Sec.~\ref{sec:ground_heisenberg}, however, since these are fine-tuned cases they will not be further discussed.

The excitation energies $\Delta E$ of the defects in the hexagons are given by
\begin{equation}
\Delta E=J_x({\mathbf S}_4+{\mathbf S}_8+{\mathbf S}_{12})^2/2\;,
\end{equation}
with
\begin{equation}
J_x=\begin{cases}J_2&\text{for the $J_1$-$J_2$-$J_{3\text{d}}$ model}\\J_{5\bigtriangleup}&\text{for the $J_1$-$J_{5\bigtriangleup}$ model}\\J_{5\bigtriangleup}+J_{5\bigtriangledown}&\text{for the $J_1$-$J_{5\bigtriangleup}$-$J_{5\bigtriangledown}$ model}\end{cases}\;,
\end{equation}
while $\Delta E$ for the defects in the triangles reads
\begin{equation}
\Delta E=J_x({\mathbf S}_2+{\mathbf S}_6+{\mathbf S}_{12})^2/2\;,
\end{equation}
with
\begin{equation}
J_x=\begin{cases}J_1-J_2&\text{for the $J_1$-$J_2$-$J_{3\text{d}}$ model}\\J_{1}&\text{for the $J_1$-$J_{5\bigtriangleup}$ model}\end{cases}\;.
\end{equation}
The energies $\Delta E$ are generally complicated functions in $u$, $v$, $w$. However, the most important property of $\Delta E$ is that it assumes continuous values between its maximum (reached for a triple fracton in the three-color models) and its minimum at $\Delta E=0$. Remarkably, the case $\Delta E=0$ does not only occur in the trivial configuration of a single homogeneous ${\mathbf q}=0$ state but also appears in non-trivial configurations with actual domains in the system. The fact that defects may have zero excitation energies and, hence, are part of the set of degenerate ground states is already included in the ground state construction of Sec.~\ref{sec:ground_heisenberg}. For example, one can construct a zero-energy defect by starting with a ${\mathbf q}=0$ state and performing the manipulations of case $(iii)$ along all lines marked $\Rightarrow\cdots\Leftarrow$ in Fig.~\ref{fig:heis_fracton}. This defect is described by the parameters $\{u=-1/2,v=-\arccos(-1/\sqrt{3}),w=-1/2\}$. We note in passing that states with more than one of these ``zero energy defects'' can be constructed. In their densest configuration they are located in the centers of {\it each} hexagon. This spin state has a twelve site unit cell given by the Star of David around the center hexagon in Fig.~\ref{fig:heis_fracton}(a) and is referred to as {\it cuboc 1} order (which appears in several contexts in kagome spin models~\cite{messio11}). Indeed, it has previously been realized that the line $J_2=J_{3\text{d}}<J_1$ of the $J_1$-$J_2$-$J_{3\text{d}}$ model as considered here, marks the phase boundary between ${\mathbf q}=0$ and {\it cuboc 1} ordered regimes~\cite{messio12,bernu13,gomez18,iida20}.
\begin{figure}
\includegraphics[width=0.99\linewidth]{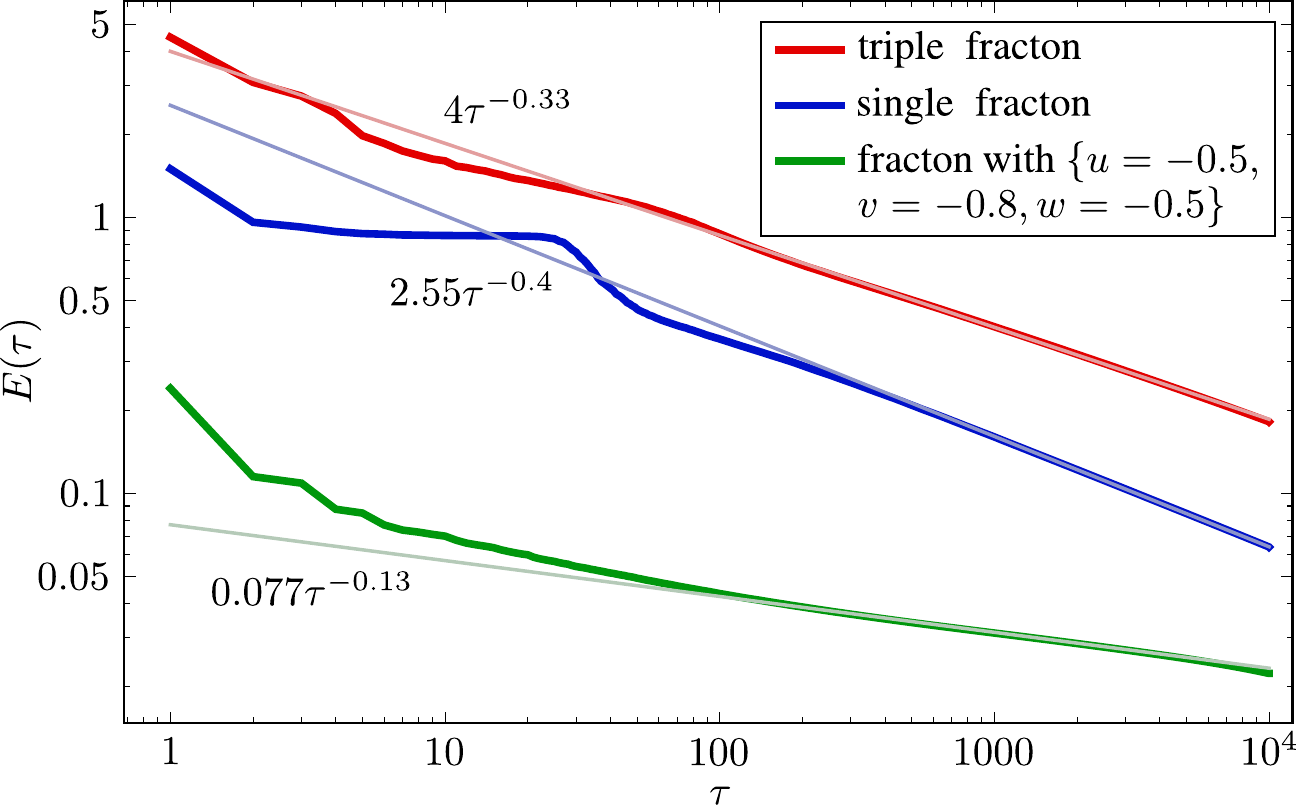}
\caption{Decay of excitation energies $E(\tau)$ of isolated defects within iterative minimization for the $J_1$-$J_{5\bigtriangleup}$ model with $J_1=J_{5\bigtriangleup}=1$ where $\tau$ is the simulation time (step count). Shown are results for three different initial defects: Triple fracton, single fracton and a low energy fracton with $\{u=-0.5,v=-0.8,w=-0.5\}$. Straight lines are power-law fits at large $\tau$ with the specific function indicated in each case. Both axes are scaled logarithmically.\label{fig:decay}}
\end{figure}

The key question is whether defects with $\Delta E>0$ are local energy minima and how they are modified when performing iterative minimization as has been done for the defects in the XY-model (see Sec.~\ref{xy_low_energy}). Indeed, the fact that single defects have a continuous energy spectrum which reaches down to zero means that they are no local minima but may decay to a ground state without passing any energy barriers. However, since such a decay amounts to changing complete sectors of ${\mathbf q}=0$ orders in Fig.~\ref{fig:heis_fracton} which involves the modification of infinitely many spins, this happens very slowly. We have tested this numerically by putting a single defect in the center of a system with open boundary conditions. The defect is relaxed by applying $10^4$ iterative minimization steps according to the procedure explained in Sec.~\ref{xy_low_energy}. As an example, we show in Fig.~\ref{fig:decay} the energy decay of three different initial states in the $J_1$-$J_{5\bigtriangleup}$ model (a triple fracton with $\Delta E=4.5J_1$, a single fracton with $\Delta E=1.5J_1$ and a fracton with $\{u=-0.5,v=-0.8,w=-0.5\}$ and $\Delta E\approx0.24J_1$). Our observations are very different from the XY model: Firstly, while in the XY model the relaxation towards a local minimum only involved a small number of spins, here, the minimization procedure approaches a ground state and involves spins across the entire system. Secondly, in contrast to the exponential energy decay for the XY model here we observe a slow relaxation process following a power-law behavior $E(\tau)\approx\tau^{-\alpha}$. Note that $\tau$ is the step count and the exponent $\alpha$ takes small values down $\alpha=0.13$ for low-energy defects.

We will return to these defects in the next section where we apply classical Monte Carlo to investigate their thermal behaviors. These studies indicate that remnants of the defects discussed here still appear in spin configurations where thermal equilibrium has not been fully reached.   

\section{Monte Carlo simulations}\label{sec:numerics}
After having discussed the properties of individual fracton-like defects we now investigate their collective and thermal behaviors via classical Monte Carlo simulations. In the following, we only treat the $J_1$-$J_{5\bigtriangleup}$ model which can be considered as the minimal model exhibiting the aforementioned fracton-like behavior. Here `minimal' refers to the fact that compared to the $J_1$-$J_{5\bigtriangleup}$-$J_{5\bigtriangledown}$ and the $J_1$-$J_2$-$J_{3\text{d}}$ models it features the smallest number of interacting bonds. It further has the property that each site contributes to three corner-sharing triangles (see~\cite{hopkinson06,isakov08,chillal20} for similar systems in 3D) and, hence, represents a generalization of nearest neighbor kagome systems (where each site is only connected to two triangles). Three versions of the $J_1$-$J_{5\bigtriangleup}$ model will be discussed below, the three state Potts, XY, and Heisenberg models, where we always set $J_1=J_{5\bigtriangleup}$.

We apply a standard Metropolis algorithm with single spin updates for a system of rhombic shape with a side length of $L$ nearest neighbor lattice spacings and untwisted periodic boundary conditions. A system characterized by the linear length $L$ then contains $N=3L^2/4$ lattice sites. In our results below, $L$ is varied between $L=40$ and $L=100$. To efficiently approach low temperatures we simulate a slow cooling process using an exponential protocol
\begin{equation}
T=T_0e^{-\gamma t}\;,
\end{equation}
where $T_0=2J_1$. Here, $t$ counts the Metropolis steps such that within one step each spin is, on average, updated once (note that we use the variable $t$ to distinguish it from the step count $\tau$ in iterative minimization). After each normal Metropolis step, we perform $10$ overrelaxation steps which helps achieving better thermalization~\cite{creutz87,kanki05,zhitomirksy08,pixley08}. For a general Heisenberg Hamiltonian $H=\sum_{i,j}J_{ij}{\mathbf S}_i\cdot{\mathbf S}_j$ an overrelaxation step amounts to randomly select a spin which is rotated by an angle of $\pi$ around the direction of the local field ${\mathbf h}_i=\sum_j J_{ij} {\mathbf S}_j$. Ensemble averages are performed with respect to $50$ independent simulation runs.
\begin{figure}
\includegraphics[width=0.95\linewidth]{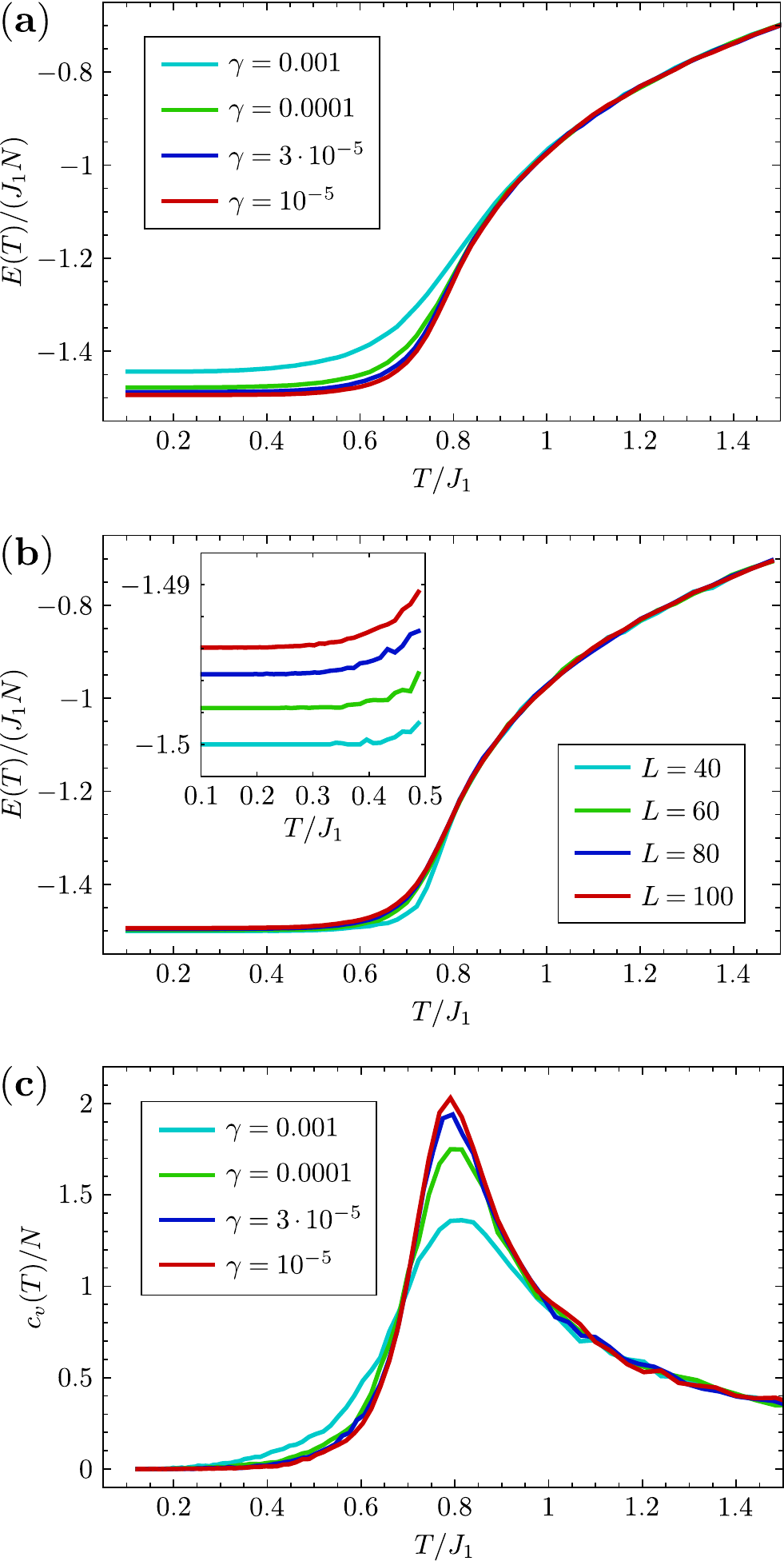}
\caption{(a) Energy per site $E(T)/(J_1 N)$ for the three-state Potts model at constant system size $L=100$ and varying cooling rates $\gamma$. (b) Energy per site $E(T)/(J_1 N)$ for the three-state Potts model at constant cooling rate $\gamma=10^{-5}$ and different system sizes $L$. The inset shows the
low temperature behavior in detail. (c) Specific heat per site $c_v(T)/N$ of the three-state Potts model for different $\gamma$ and constant $L=100$.\label{fig:potts_e}}
\end{figure}
\begin{figure}
\includegraphics[width=0.99\linewidth]{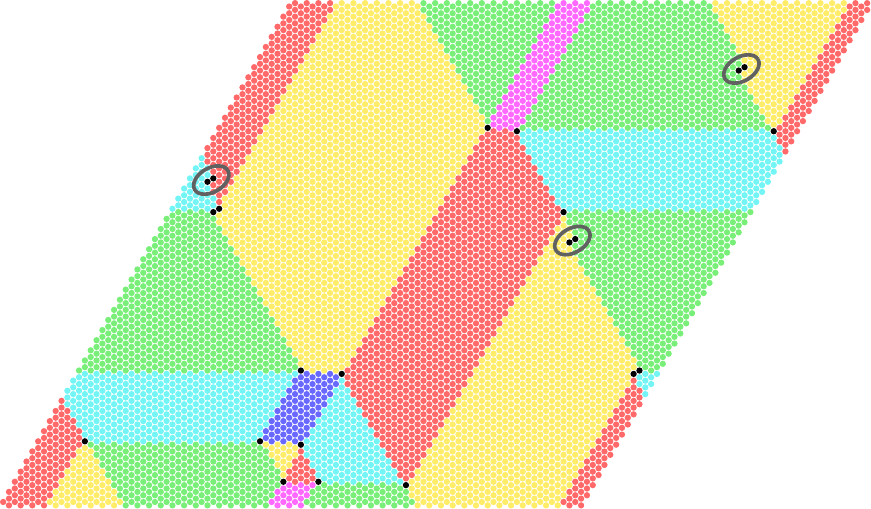}
\caption{Typical output of a spin configuration from Monte Carlo for the three state Potts model. The simulation has been performed for a system size of $L=100$, at $T=0.1J_1$ and cooling rate $\gamma=10^{-5}$. Each point represents a $J_1$ or $J_{5\bigtriangleup}$ triangle of the original lattice where the colors of the points indicate the six possible local ${\mathbf q}=0$ orders, see Fig.~\ref{fig:potts_ground_states}. Black points denote defect triangles and gray circles highlight the lineons.\label{fig:potts_spins}}
\end{figure}

\subsection{Three-state Potts model}
\subsubsection{Internal energy, specific heat and low temperature spin configurations}
We start discussing the internal energy per site $E(T)/N$ as a function of $T$ for different cooling rates $\gamma$ [Fig.~\ref{fig:potts_e}(a)] and for different system sizes $L$ [Fig.~\ref{fig:potts_e}(b)]. Our data in Fig.~\ref{fig:potts_e}(a) indicates that small $\gamma$ are essential to approach the exact ground state energy of $E_\text{exact}(T=0)/N=-1.5J_1$ at small temperatures. Indeed, even for the smallest $\gamma=10^{-5}$ the system may get trapped in a local minimum below $T\approx0.5J_1$ where $E(T)$ is a flat line with a small offset compared to the ground state energy. The dependence on the system size $L$ is shown in Fig.~\ref{fig:potts_e}(b) where $\gamma$ is kept constant at $\gamma=10^{-5}$. While at large $T$ results are well converged in $L$, a characteristic observation at small $T$ is that $E(t)$ slightly {\it increases} with $L$ [see inset of Fig.~\ref{fig:potts_e}(b)] indicating that in larger systems the defect density is higher. This behavior can be understood from the fact that moving a fracton requires flipping spins across the entire system and, consequently, the probability for fractons to recombine and annihilate is larger for small systems.

It might seem tempting to compare the simulated $E(T)$ to an exact solution as is possible for other kinetically constrained models~\cite{newman99,garrahan00,ritort03,jack05,lipowski97,garrahan02}. Indeed, many of such models have a dual description which allows for a direct mapping between independent defect variables and spin states such that the thermodynamics becomes trivial and an exact solution may be easily formulated. In our case, however, we could not identify such a dual representation and, consequently, defects cannot be considered as independent. For the $J_1$-$J_2$-$J_{3\text{d}}$ model this is immediately obvious since for two triple defects in adjacent $J_1$ triangles, the nearby hexagons must, likewise, carry defects. We can, however, not exclude the possibility that an effective dual representation exists in a suitably defined low-energy subspace.

In Fig.~\ref{fig:potts_e}(c) we present the system's specific heat for varying $\gamma$ at constant $L=100$, calculated by differentiating $E(T)$. Due to the gapped nature of spin excitations we find activated behavior at low $T$ and zero slope $\frac{d c_v(T)}{dT}$ for $T\rightarrow0$. An obvious feature is a pronounced peak at $T\approx0.8J_1$. Below, we will argue that this peak is associated with a crossover into a spin glass-like regime where the system becomes non-ergodic.

A typical spin configuration in a local minimum obtained for $T=0.1J_1$ and $\gamma=10^{-5}$ is shown in Fig.~\ref{fig:potts_spins} where each $J_1$ and $J_{5\bigtriangleup}$ triangle is represented by a point (which together form a triangular lattice). If a $J_1$ or $J_{5\bigtriangleup}$ triangle is in a local ground state (i.e., the three-color constraint is fulfilled) the color of the point encodes the six possible configurations. A triangle violating the color constraint is illustrated as a black point. As can be seen, the system exhibits a patchwork of different ${\mathbf q}=0$ order domains with single fractons at positions where four domain walls meet. Note that triple fractons cost too much energy to be observed at $T=0.1J_1$. The system also features various lineons consisting of two single fractons, indicated by gray circles.
\begin{figure*}
\includegraphics[width=0.99\linewidth]{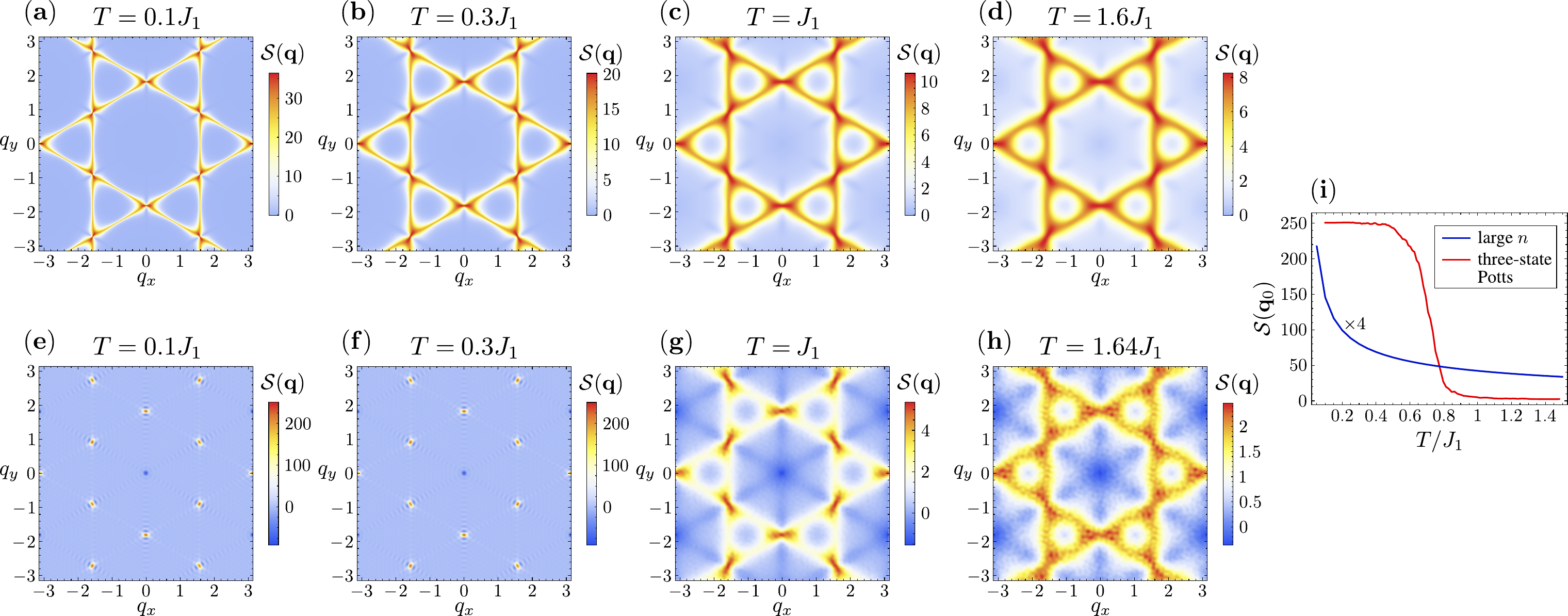}
\caption{Spin structure factor of the $J_1$-$J_{5\bigtriangleup}$ model for different temperatures $T$. Top row [(a)-(d)]: Spin structure factor in large $n$ approximation. Bottom row [(e), (f)]: Spin structure factor of the three-state Potts model from classical Monte Carlo for $\gamma=10^{-5}$ and $L=100$. The lattice constant of the kagome lattice has been set to unity for the definition of ${\mathbf q}$. (i) Spin structure factor $\mathcal{S}({\mathbf q}_0)$ at the ${\mathbf q}=0$ order position ${\mathbf q}_0=(0,\pi/\sqrt{3})$ as a function of temperature in large $n$ approximation (blue line, rescaled by a factor of $4$ for better visibility) and from Monte Carlo simulations in the three-state Potts model (red line). Note that the spin structure factor in the large $n$ approximation is only calculated for one spin component.} \label{fig:potts_spin_structure}
\end{figure*}

\subsubsection{Spin structure factor}
Next, we discuss the equal-time spin structure factor
\begin{equation}
\mathcal{S}({\mathbf q})=\frac{1}{N}\sum_{i,j}e^{-i{\mathbf q}\cdot({\mathbf r}_i-{\mathbf r}_j)}\langle{\mathbf S}_i\cdot{\mathbf S}_j\rangle\;,\label{ssf}
\end{equation}
where ${\mathbf r}_i$ is the position of site $i$ and $\langle\ldots\rangle$ averages over independent simulation runs. For better interpretation of the Monte Carlo data, we first present results of an $O(n)$ approximation where the spins ${\mathbf S}_i$ are generalized to $n$-component vectors ${\mathbf S}_i=(S_i^1,S_i^2,\ldots,S_i^n)$ subject to local length constraints $\sum_{\mu=1}^n (S_i^\mu)^2=1$. The system can be treated exactly in the limit of large $n$~\cite{garanin99,moshe03,isakov04}, where thermal fluctuations are correctly accounted for even if ergodicity is lost. While the generalization from a three state Potts model (which is a discretized version of an $n=2$ system) to $n\rightarrow\infty$ may appear drastic, previous results from pyrochlore magnets indicate that the large $n$ limit provides an excellent approximation for models with small $n$~\cite{isakov04}. The large $n$ approximation uses the eigenvalues $\epsilon_\alpha({\mathbf q})$ and corresponding eigenvectors $\psi_\alpha({\mathbf q})$ (where $\alpha=1,2,3$) of the coupling matrix
\begin{equation}
J_{\alpha\beta}({\mathbf q})=\sum_b e^{-i{\mathbf q}\cdot({\mathbf r}_{0\alpha}-{\mathbf r}_{b\beta})}J_{0\alpha b\beta}\;.
\end{equation}
Here, we have split up the site index $i$ into two indices $i\rightarrow (a,\alpha)$ where $a$ denotes the unit cell and $\alpha$ enumerates sites within a unit cell. In this notation couplings and site positions read $J_{ij}\rightarrow J_{a\alpha b\beta}$ and ${\mathbf r}_i\rightarrow{\mathbf r}_{a\alpha}$. The spin structure factor is then obtained via
\begin{equation}
\mathcal{S}({\mathbf q})=\sum_{\alpha,\beta,\gamma=1}^3\frac{[\psi_\beta({\mathbf q})]_\alpha[\psi^*_\beta({\mathbf q})]_\gamma}{\frac{\epsilon_\beta({\mathbf q})}{T}+\lambda}\;,
\end{equation}
where $\lambda$ is a Lagrange multiplier enforcing spin normalization, which is determined from the condition
\begin{equation}
\frac{1}{N}\sum_{{\mathbf q}\in\text{BZ}}\sum_{\alpha=1}^3\left[\frac{\epsilon_\alpha({\mathbf q})}{T}+\lambda\right]=1\;.
\end{equation}
In Fig.~\ref{fig:potts_spin_structure}(a)-(d) we show the spin structure factor of the $J_1$-$J_{5\bigtriangleup}$ model at large $n$ for various temperatures $T$. At small $T$ the response is entirely distributed along streaks in momentum space (forming again a kagome lattice) which is a direct consequence of the system's subextensive degeneracy. The three directions of the streaks corresponds to the three directions of lines along which the spins of a ${\mathbf q}=0$ state can be swapped to form a new ground state. With increasing $T$ the streaks broaden and pinch point-like patterns become visible at their intersections. Similar features at the same momenta are well known from nearest neighbor kagome antiferromagnets which are characterized by an {\it extensive} ground state degeneracy. These similarities indicate that at finite temperatures, the system may explore parts of this extensive manifold of states. It can further be seen that the broadening of streaks occurs in a continuous manner without any noticeable abrupt changes.

Turning to the Monte Carlo results [see Fig.~\ref{fig:potts_spin_structure}(e)-(h)] we observe qualitative agreement with the large $n$ approximation at sufficiently large temperatures $T\gtrsim J_1$ where broadened streaks dominate the magnetic response. The intactness of these streaks indicates that the system undergoes the characteristic fluctuations of color swaps along lines of alternating colors. Lowering the temperature below $T\approx0.8J_1$ (which is the peak position in the specific heat), the signal shows a sudden rearrangement not seen in large $n$, where pronounced peaks at ${\mathbf q}=0$ order positions (located at the intersections of the streaks) emerge. The absence of streaks in momentum space indicates that the system can no longer realize the aforementioned line-like fluctuations since any intersection of a flipped two-color line with a domain wall creates defects associated with an excitation energy. The system, hence, gets arrested in states with patch-like patterns of different ${\mathbf q}=0$ domains as shown in Fig.~\ref{fig:potts_spins}. To further investigate the thermal behavior of domains it is instructive to plot the spin structure factor at the ${\mathbf q}=0$ order positions [e.g. at ${\mathbf q}_0=(0,\pi/\sqrt{3})$] as a function of temperature, see Fig.~\ref{fig:potts_spin_structure}(i). In an ideal ${\mathbf q}=0$ state each site contributes $4/9$ to $\mathcal{S}({\mathbf q}_0)$ while on length scales larger than the typical domain size the correlations decay to zero. Hence, the spin structure factor at the ${\mathbf q}=0$ order positions provides a direct measure of domain sizes. Figure~\ref{fig:potts_spin_structure}(i) reveals that below the steep increase at $T\approx0.8J_1$ this quantity further grows, indicating a limited mobility of domain walls, but eventually shows a plateau-like behavior at small temperatures [this is in contrast to the smooth increase of $\mathcal{S}({\mathbf q}_0)$ in large $n$ approximation, see blue line in Fig.~\ref{fig:potts_spin_structure}(i)]. Note that the plateau value is still significantly smaller than the maximal possible value of $716$ due to the system's finite size and, hence, the plateau is no finite-size effect. Obviously, in this low temperature regime the system gets stuck in local minima. We, therefore, conclude that the system undergoes a glass-like transition at $T\approx0.8J_1$.

For a 2D system without any continuous degrees of freedom, a finite-temperature transition to a long-range magnetically ordered state is generally possible which in our case would be associated with a thermal order-by-disorder selection. However, the system does not display this behavior. The low-temperature spin configurations are given by a dilute gas of single defects (which are the lowest energy excitations) as shown in Fig.~\ref{fig:potts_spins}. These defects are connected by a network of domain walls, where each wall disrupts the ${\mathbf q}=0$ correlations. Indeed, already one single defect in an otherwise defect-free system has domains walls sticking out in four directions which is sufficient to destroy long-range ${\mathbf q}=0$ order.
\begin{figure}
\includegraphics[width=0.9\linewidth]{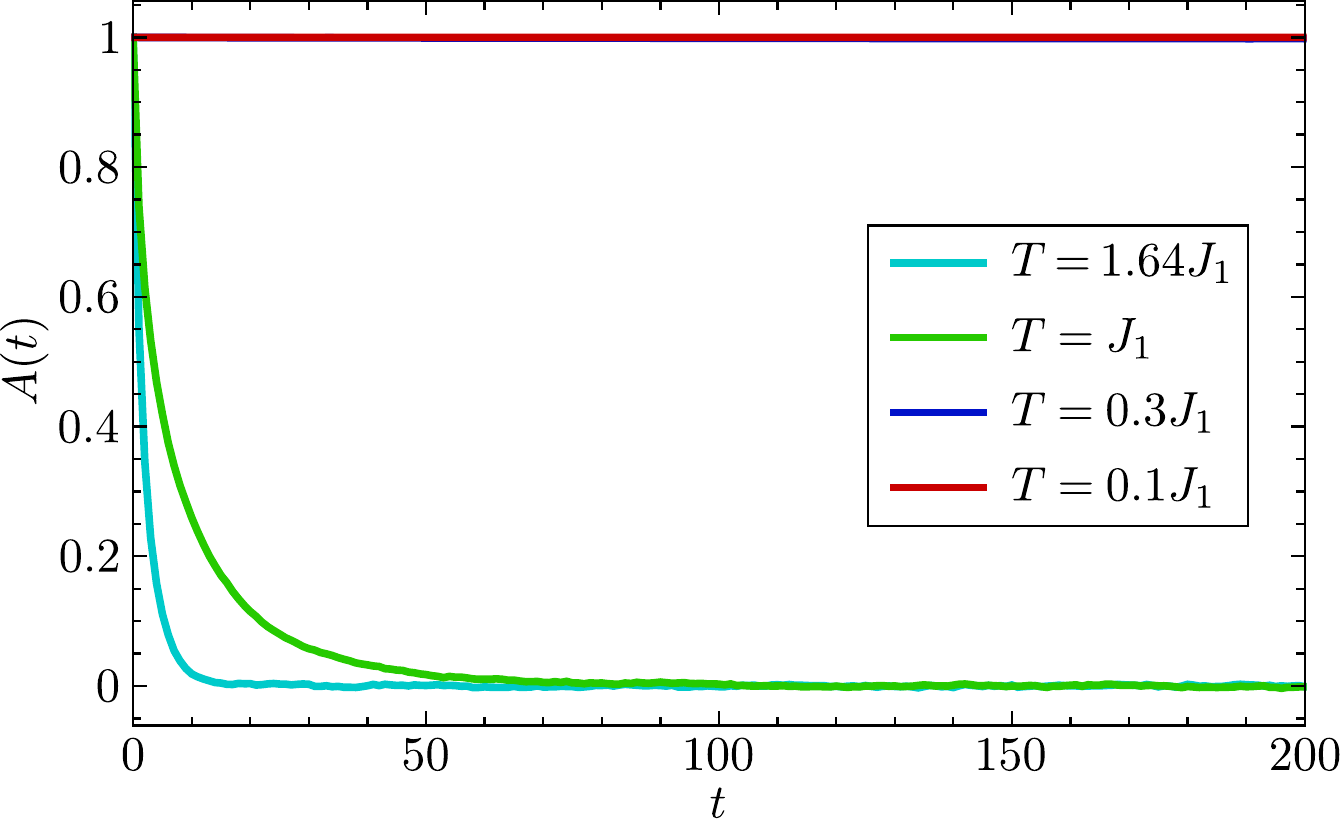}
\caption{Autocorrelation function $A(t)$ [see Eq.~(\ref{auto_correlation})] of the three-state Potts model for system size $L=100$ and different temperatures $T$. The initial equilibration prior to the measurement of $A(t)$ has been performed with a cooling rate of $\gamma=10^{-5}$. Note that the blue and red curves lie almost on top of each other. \label{fig:potts_auto_correlation}}
\end{figure}

\subsubsection{Autocorrelation function}
To further substantiate the glassy dynamics at small temperatures, we discuss the system's autocorrelation function
\begin{equation}
A(t)=\frac{1}{N}\sum_i\langle {\mathbf S}_i(t_0)\cdot{\mathbf S}_i(t_0+t)\rangle\label{auto_correlation} 
\end{equation}
for varying temperatures. Again, $\langle\ldots\rangle$ denotes an averaging over independent simulation runs. The time argument of the spin variables denotes the Monte Carlo step. The initial time $t_0$ is given by the Monte Carlo step at which our exponential cooling protocol reaches the desired temperature $T$. Since we use the smallest cooling rate $\gamma=10^{-5}$ for the initial equilibration, the time $t_0$ corresponds to a ``long waiting time'' such that for large $t$ the autocorrelation function $A(t)$ approximates the Edwards-Anderson parameter~\cite{edwards75} for spin glasses $q_\text{EA}$ defined as
\begin{equation}
q_\text{EA}=\lim_{t\rightarrow\infty}\lim_{t_0\rightarrow\infty}\lim_{L\rightarrow\infty}\frac{1}{N}\sum_i\langle {\mathbf S}_i(t_0)\cdot{\mathbf S}_i(t_0+t)\rangle\;.\label{qea}
\end{equation}
After the time $t_0$, the temperature is kept constant and the data sampling for $A(t)$ starts.

The results in Fig.~\ref{fig:potts_auto_correlation} confirm the expected behavior. For temperatures $T\gtrsim J_1$ the autocorrelation function $A(t)$ quickly decays to zero showing that the system loses its memory about the initial state. On the other hand, in the low temperature regime $T\lesssim0.3J_1$ practically no evolution in $t$ is observed and the autocorrelation functions remain close to one.

It is worth emphasizing that even though the system shows signatures of glassy behavior, it still differs from conventional spin glasses. According to common understanding a spin glass exhibits a distribution of energy barriers whose heights scale as a power law in $L$. In our case, however, removing an isolated fracton in an otherwise defect-free system is associated with a constant energy barrier (on the order of $J_1$) independent of system size. Rather, the slow dynamics at small $T$ is due to the fact that the minimal number of spin flips required to remove such a defect scales with a power law in $L$. 
\begin{figure}
\includegraphics[width=0.95\linewidth]{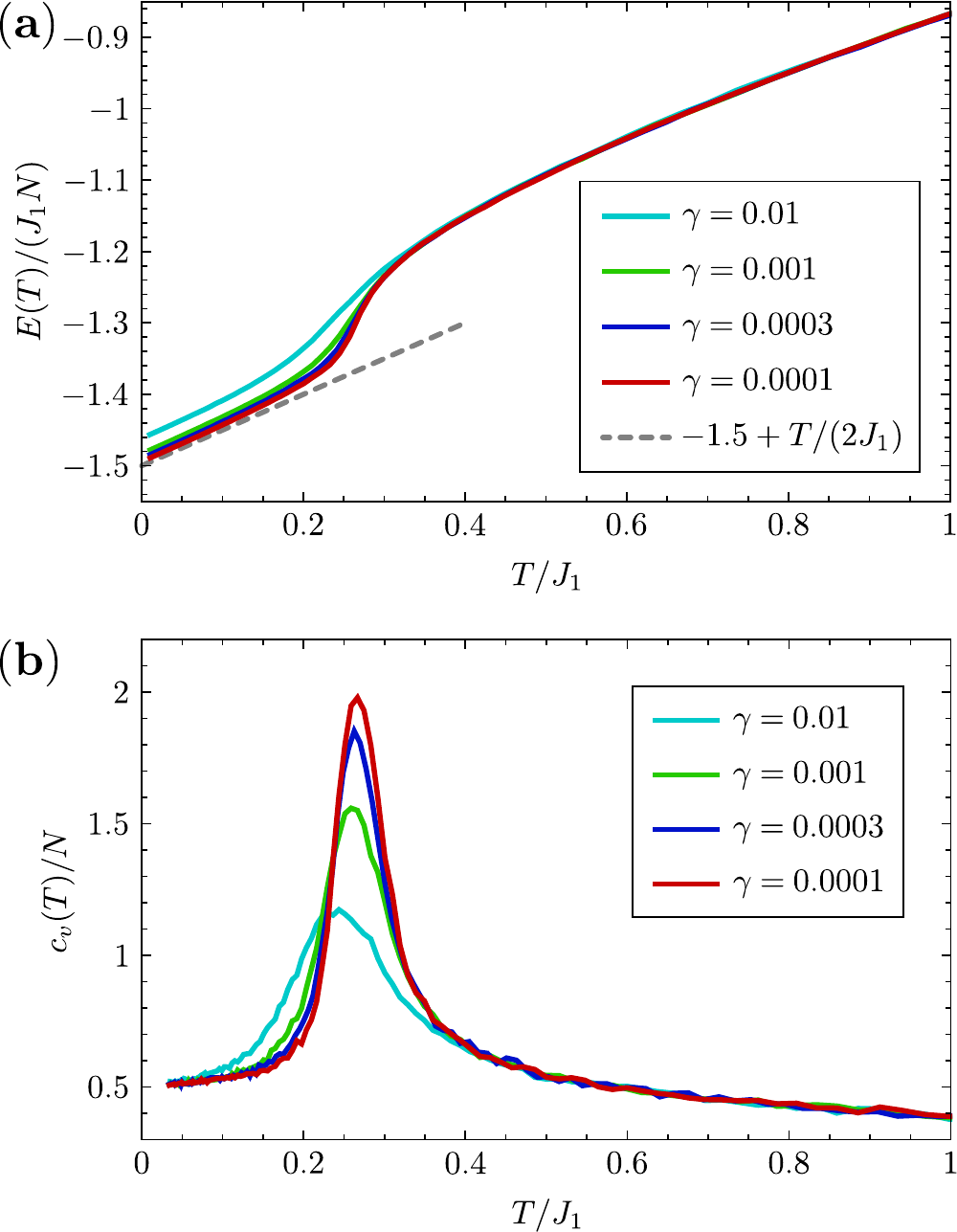}
\caption{(a) Energy per site $E(T)/(J_1N)$ for the XY $J_1$-$J_{5\bigtriangleup}$ model with $L=100$ and varying $\gamma$. At small temperatures, the Monte Carlo data approximately follows $E(T)/N=-1.5J_1+T/2$, as indicated by the dashed line. (b) Specific heat $c_v(T)/N$ for $L=100$ and varying $\gamma$ obtained by differentiating $E(T)$.  \label{fig:xy_e}}
\end{figure}
\begin{figure*}
\includegraphics[width=0.99\linewidth]{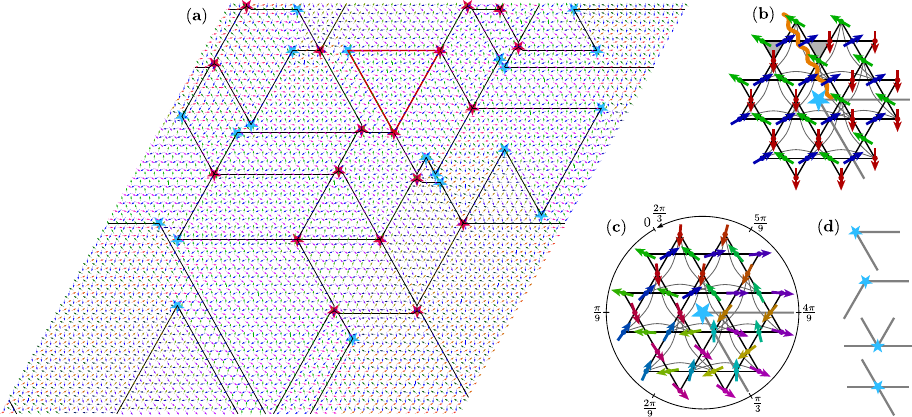}
\caption{(a) Real space spin configuration in the XY $J_1$-$J_{5\bigtriangleup}$ model at $T=0.01J_1$ obtained for $\gamma=0.0001$. Domain walls are illustrated by black lines defined via antiparallel chiralities $\kappa$ [see Eq.~\ref{chirality}] on neighboring $J_1$ triangles. Fractons (fractional vortices) are indicated by magenta (blue) stars. The triangular domain walls highlighted by red lines in the upper part of the figure show a typical pattern where the positions of one vortex and two fractons are fixed with respect to each other. (b) Construction of a fractional vortex starting from a three-color state with a kinked domain wall (thick gray line), see text for details. The blue star indicates the vortex core and the wavy orange line is a branch cut of defect triangles. (c) Fractional vortex obtained after rotating the spins in (b) by an angle specified on the outer circle. (d) Alternative domain wall configurations of fractional vortices.  \label{fig:fractional_vortex}}
\end{figure*}

\subsection{XY model}\label{numerics_xy}
\subsubsection{Internal energy, specific heat and low temperature spin configurations}
Next, we investigate thermodynamic properties of the XY $J_1$-$J_{5\bigtriangleup}$ model. The system's internal energy per site $E(T)/N$ from classical Monte Carlo for various $\gamma$ and fixed $L=100$ is shown in Fig.~\ref{fig:xy_e}(a). In comparison to the three-state Potts model, convergence in the cooling rate is much better, i.e., $\gamma$ can be chosen significantly larger. This is due to the fact that the continuous spin degrees of freedom facilitate the system's escape from local energy minima. Similarly, convergence with respect to system size $L$ (not shown here) is also found to be much better. At low temperatures the internal energy is well approximated by $E(T)/N=-1.5J_1+T/2$ where $E_\text{exact}(T=0)/N=-1.5J_1$ is the exact ground state energy and the term $T/2$ is the expected contribution from a single quadratic mode per site. The system's specific heat $c_v(T)$ obtained from differentiating $E(T)$ is plotted in Fig.~\ref{fig:xy_e}(b). As will be discussed further below, a pronounced peak at $T\approx0.27J_1$ is again associated with a crossover into a glassy phase.

A typical low temperature real space spin configuration is shown in Fig.~\ref{fig:fractional_vortex}(a). Apart from the individual spin orientations, the figure also highlights the domains walls between different local ${\mathbf q}$ orders which we define via the spin chiralities $\kappa$ on $J_1$-triangles
\begin{equation}
\kappa={\mathbf S}_1\times{\mathbf S}_2+{\mathbf S}_2\times{\mathbf S}_3+{\mathbf S}_3\times{\mathbf S}_1\;.\label{chirality}
\end{equation}
Here, ${\mathbf S}_1$, ${\mathbf S}_2$, ${\mathbf S}_3$ are the three spins in a (upward- or downward pointing) $J_1$-triangle ordered in a counterclockwise manner. If two neighboring triangles have parallel (antiparallel) chiralities $\kappa$ they belong to the same domain (are separated by a domain wall passing through the shared spin). As in the three-state Potts model, single fractons are clearly observed via their characteristic structure of crossing domain walls, indicated by magenta stars. Interestingly, however, one also finds kinks in the domain walls enclosing angles of $\pi/3$ or $2\pi/3$ (blue stars) which are forbidden in a three color model. We argue in the following, that these spin configurations are fractional vortices known from classical nearest neighbor XY kagome antiferromagnets~\cite{rzchowski97,korshunov02,zhitomirksy08}.

To start with, we illustrate in Fig.~\ref{fig:fractional_vortex}(b) an attempt to construct a three-color spin state with a domain wall exhibiting a $\pi/3$ kink (thick gray line). Obviously, the restriction to only three spin orientations creates mismatches along a branch cut (orange wavy line) where defect triangles are unavoidable. Comparing the two gray shaded triangles in Fig.~\ref{fig:fractional_vortex}(b) reveals that on both sides of the branch cut the spin configurations differ by a cyclic permutation of the three colors (which is equivalent to a $2\pi/3$ spin rotation). Hence, when allowing for {\it continuous} in-plane spin orientations, this branch cut can be removed by a rotation of all spins by an angle $\varphi\in[0,2\pi/3)$ which varies continuously when encircling the kink [such rotation angles are specified on the outer circle in Fig.~\ref{fig:fractional_vortex}(c)]. The resulting spin configuration in Fig.~\ref{fig:fractional_vortex}(c) can be considered as a $1/3$ fraction of a conventional integer vortex since the local trio of spins rotates by an angle of $2\pi/3$ when moving around the core. There are several other possible domain wall structures for fractional vortices where either two or four domain walls emanate from the core, see Fig.~\ref{fig:fractional_vortex}(d). Note that vortices with an odd number of emanating domain walls are not possible. This is because when trying to construct such vortices in a three color model, the states on both sides of the branch cut differ by an {\it odd} permutation. Such a mismatch cannot be removed by a continuous twist but requires a mirror operation in spin space (which is equivalent to inserting another domain wall).
\begin{figure}
\includegraphics[width=0.99\linewidth]{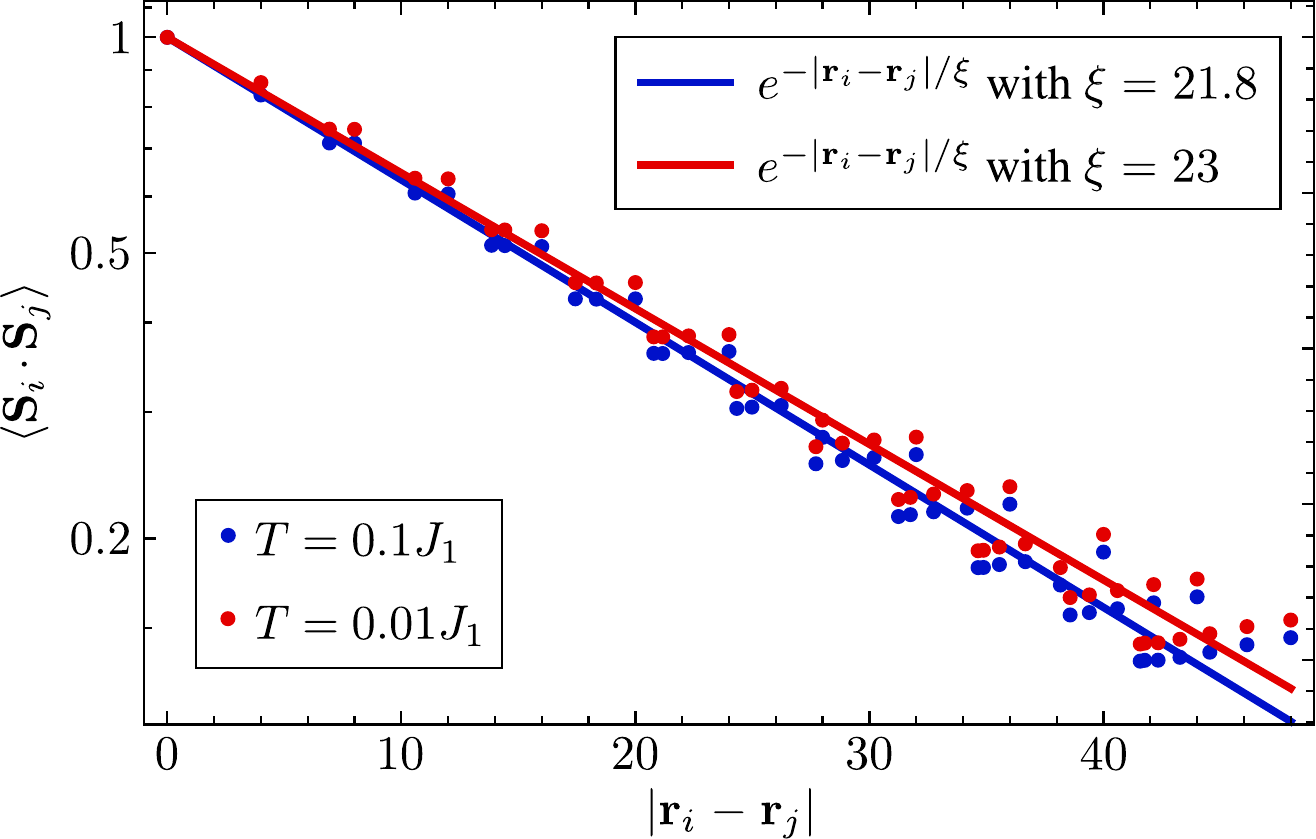}
\caption{Real-space spin correlations $\langle {\mathbf S}_i\cdot{\mathbf S}_j\rangle$ of the XY $J_1$-$J_{5\bigtriangleup}$ model as a function of the distance $|{\mathbf r}_i-{\mathbf r}_j|$ between spins for temperatures $T=0.01$ and $T=0.1$ using $\gamma=0.0001$ and $L=100$. To avoid sublattice effects, only sites on the same kagome-sublattice have been considered. Lines are exponential fits of the data.   \label{fig:xy_correlations}}
\end{figure}

Fractional vortices are well known in nearest neighbor XY kagome antiferromagnets where they undergo a usual Kosterlitz-Thouless binding/unbinding transition. The associated transition temperatures are, however, substantially reduced compared to models with integer vortices~\cite{rzchowski97,korshunov02}. As discussed below, in our systems the Kosterlitz-Thouless transition even appears completely suppressed. A first obvious difference compared to more conventional XY magnets is that fractional vortices cannot be placed everywhere in the system without creating additional defects. Particularly, since fractional vortices exhibit domain walls, their real space positions highly depend on the fracton positions. A typical configuration is shown in the upper part of Fig.~\ref{fig:fractional_vortex}(a) where a triangle of domain walls is highlighted by red lines. In this spin arrangement two single fractons and a fractional vortex of the type of Fig.~\ref{fig:fractional_vortex}(c) are forced to form an equilateral triangle. [In contrast to Fig.~\ref{fig:fractional_vortex}(c), however, the $2\pi/3$ twist is not evenly distributed around the vortex core but is only found in the $(2\pi-\pi/3)$-segment outside the kink. Inside the $\pi/3$ segment a homogeneous ${\mathbf q}=0$ state is realized.] Another difference compared to conventional XY magnets is that depending on the precise arrangement of domain walls, a fractional vortex-antivortex pair does not necessarily decay to a ground state when merged, but may result in a fracton. As a consequence of these restrictions, thermal fluctuations of vortex states which drive the Kosterlitz-Thouless transition in more conventional XY magnets~\cite{kosterlitz73,berezinskii71,berezinskii72} are strongly suppressed. Indeed, our systems do not display the characteristic properties of the quasi long-range ordered state below a Kosterlitz-Thouless transition. Firstly, even at the lowest simulated temperatures a complete binding of vortices into short-distance vortex/antivortex pairs is not observed in our numerical outputs. Secondly, the spin correlations $\langle {\mathbf S}_i\cdot{\mathbf S}_j\rangle$ show a clear exponential decay as a function of the distance between sites $i$ and $j$ down to $T=0.01J_1$, see Fig.~\ref{fig:xy_correlations}. This is opposed to the power-law correlations and short-distance vortex-antivortex pairs observed in conventional XY magnets below the Kosterlitz-Thouless transition. It is worth emphasizing, however, that due to slow thermalization our low-temperature Monte-Carlo data should be interpreted with caution. Hence, we cannot generally exclude the possibility that a perfectly thermalized ensemble of states would show a Kosterlitz-Thouless transition at small temperatures.

\begin{figure}
\includegraphics[width=0.9\linewidth]{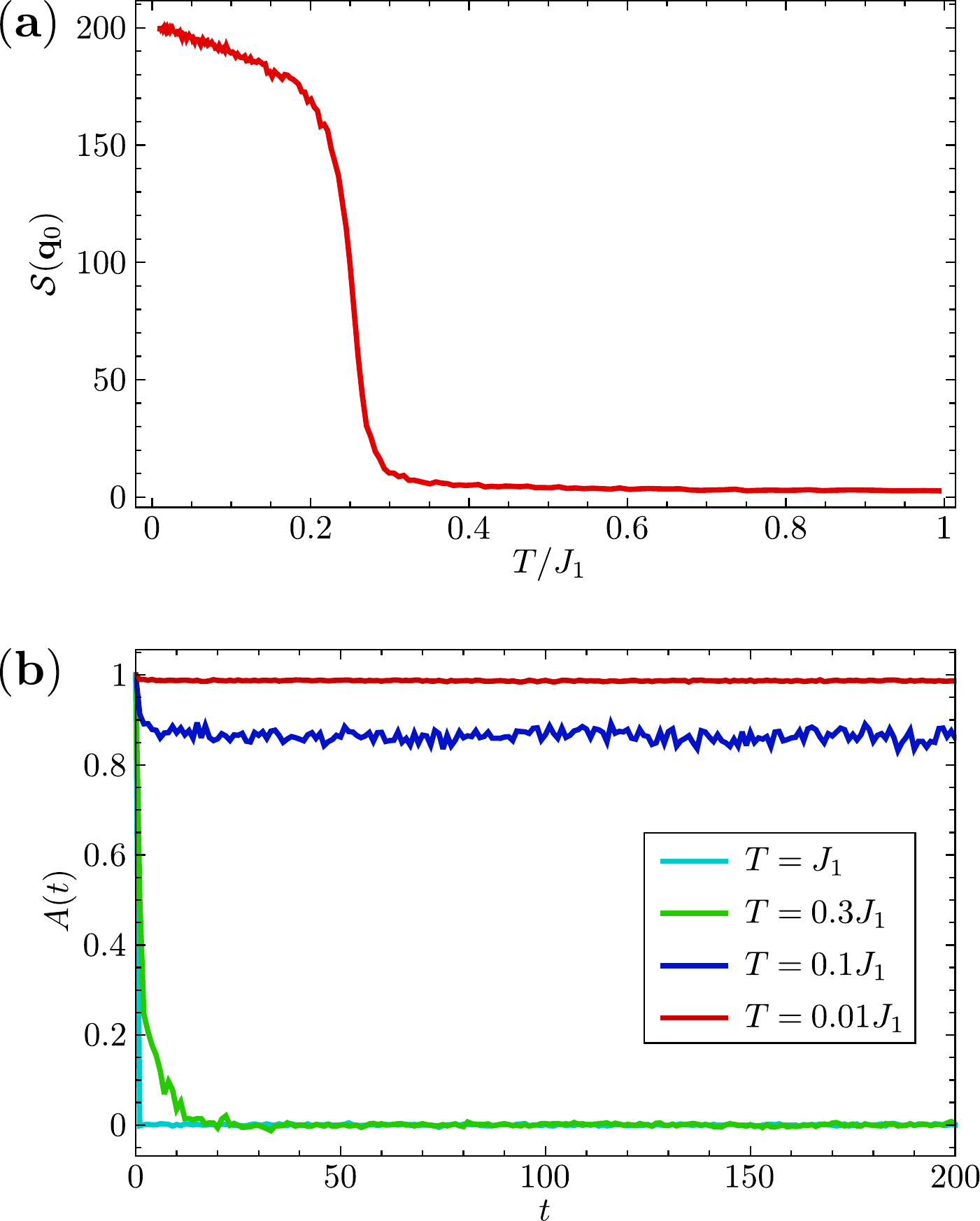}
\caption{(a) Spin structure factor $\mathcal{S}({\mathbf q}_0)$ of the XY $J_1$-$J_{5\bigtriangleup}$ model at the ${\mathbf q}=0$ order position ${\mathbf q}_0=(0,\pi/\sqrt{3})$ as a function of temperature for $\gamma=0.0001$ and $L=100$. (b) Autocorrelation function of the XY $J_1$-$J_{5\bigtriangleup}$ model [see Eq.~(\ref{auto_correlation})] at different temperatures $T$ for $L=100$. The initial equilibration has been performed with a cooling rate of $\gamma=0.0001$.
\label{fig:peak_height_XY}}
\end{figure}

\subsubsection{Spin structure factor and autocorrelation function}
In this subsection we demonstrate based on the spin structure factor and the autocorrelation function that despite the larger configuration space of the spins and the occurrence of vortices, the system still behaves glassy at small temperatures.

The spin structure factors of the XY $J_1$-$J_{5\bigtriangleup}$ model and the three-state Potts model (presented in Fig.~\ref{fig:potts_spin_structure}) are qualitatively similar. Particularly, streaks of strong signal forming a kagome lattice are found at large temperatures while sharp peaks at ${\mathbf q}=0$ order positions dominate the response at small $T$. To capture this behavior it is sufficient to plot the spin structure factor at these peak positions, see Fig.~\ref{fig:peak_height_XY}(a). Similar to the Potts model, this quantity shows a sharp increase (approximately at $T\approx0.27J_1$) which matches the temperature of the peak in the specific heat. Compared to the three state Potts model, however, this crossover is significantly reduced in temperature. Below $T\approx0.2J_1$, the peaks only show a moderate increase, indicating that the system enters a glassy phase with reduced domain wall motion.

The system's glassiness is also reflected in the autocorrelation function as defined in Eq.~(\ref{auto_correlation}) and plotted in Fig.~\ref{fig:peak_height_XY}(b). We note that in these results we explicitly eliminated a possible global drift, i.e. an overall spin rotation as a function of Monte Carlo time. Such effects lead to a spurious decay of the autocorrelation function especially at small system sizes $L$. To remove this behavior we rotated the spin state at time $t_0+t$ into a frame where for a given site $i$ the spin directions ${\mathbf S}_i(t_0)$ and ${\mathbf S}_i(t_0+t)$ are identical. Effectively, this allows one to reduce finite size effects~\cite{berthier04}. As shown in Fig.~\ref{fig:peak_height_XY}(b) the temporal behavior of the autocorrelation function is clearly different in the high temperature regime $T\gtrsim0.3$, where a rapid decay to zero is found, and at low temperatures $T\lesssim0.1$, where correlations remain finite in the long-time limit. 
\begin{figure}
\includegraphics[width=0.95\linewidth]{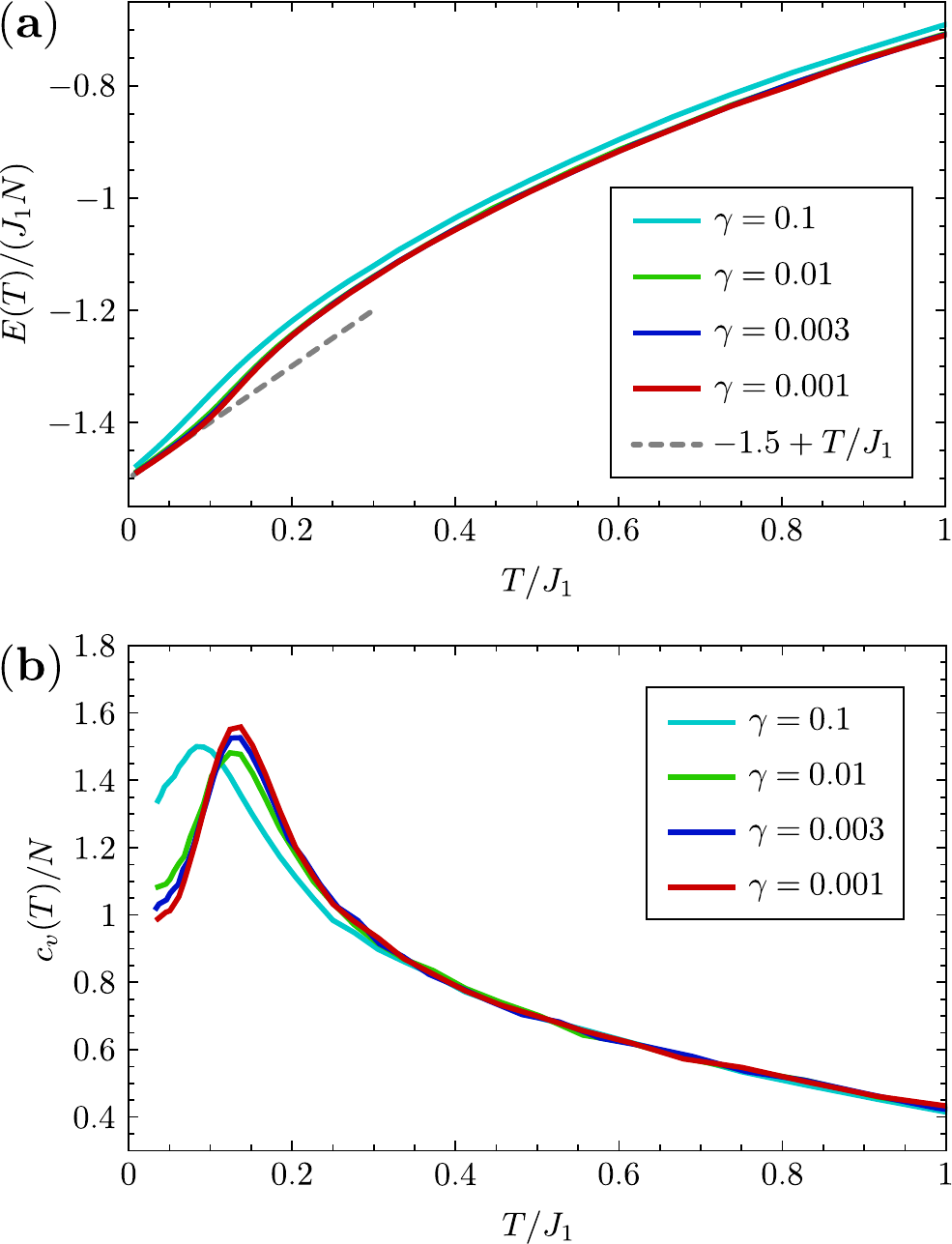}
\caption{(a) Energy per site $E(T)/(J_1N)$ for the Heisenberg $J_1$-$J_{5\bigtriangleup}$ model with $L=100$ and varying $\gamma$. The Monte Carlo data approximately behaves as $E(T)/N=-1.5J_1+T$ at small temperatures, as indicated by the dashed line. (b) Specific heat $c_v(T)/N$ for $L=100$ and varying $\gamma$ obtained by differentiating $E(T)$.     
\label{fig:heis_e}}
\end{figure}
\begin{figure}
\includegraphics[width=0.9\linewidth]{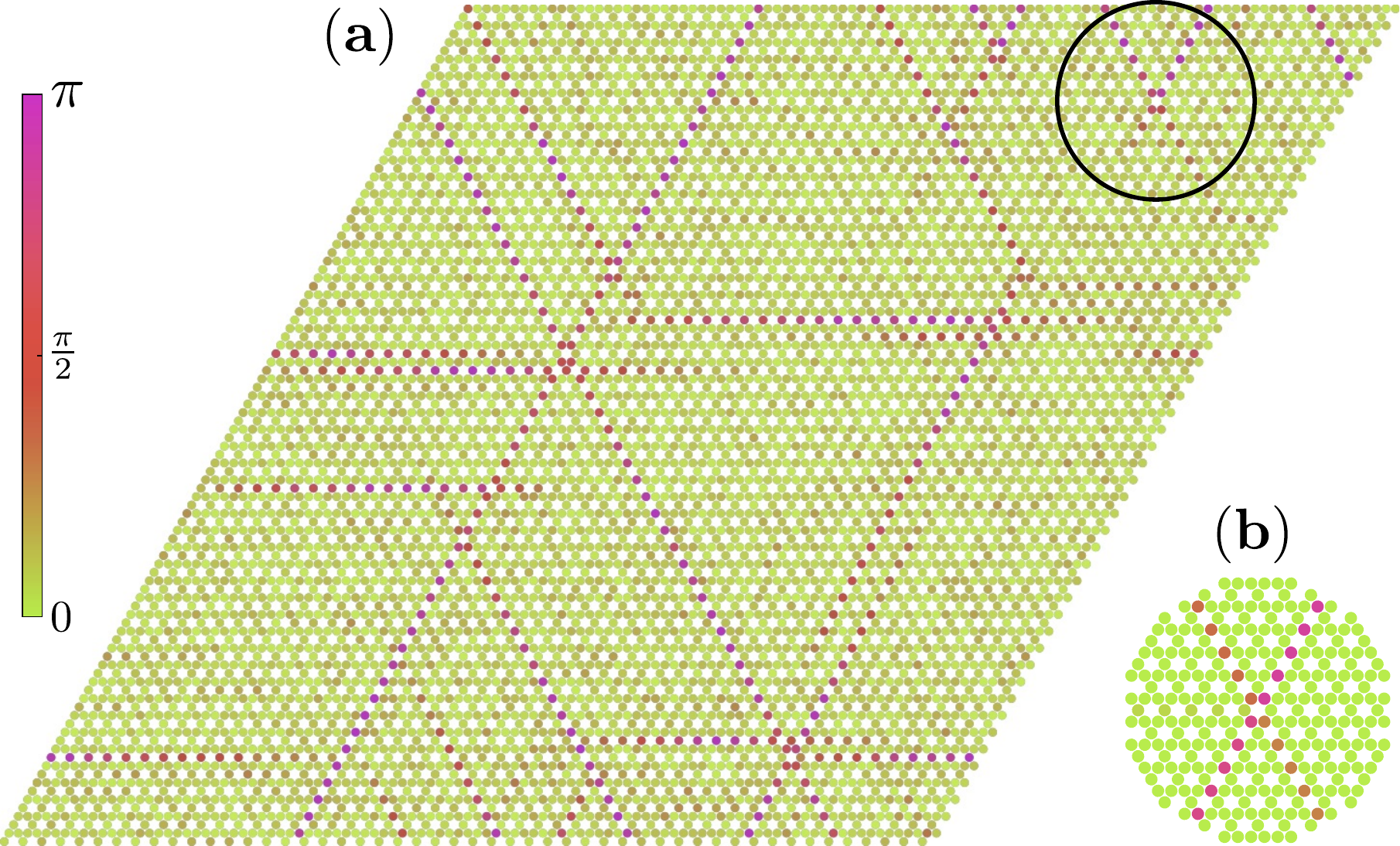}
\caption{(a) Real space spin configuration of a numerical outcome obtained for the Heisenberg $J_1$-$J_{5\bigtriangleup}$ model at $T=0.01J_1$, $L=100$, and $\gamma=0.001$. For each site, the angle between the spin chiralities [Eq.~(\ref{chirality})] on the two adjacent $J_1$ triangles is plotted. The black circle highlights a typical decaying defect state. (b) For comparison, an intact isolated defect in a single hexagon with $u=1$, $v=0.57$, and $w=-0.8$ (see Sec.~\ref{sec:ground_heisenberg}) is illustrated, using the same plotting scheme as in (a). \label{fig:heis_fracton_example}}
\end{figure}
\subsection{Heisenberg model}
\subsubsection{Internal energy, specific heat and low temperature spin configurations}
In this section we discuss the thermal behavior of the Heisenberg $J_1$-$J_{5\bigtriangleup}$ model with a particular focus on the fate of defect states. The trend towards faster equilibration when going from the three-state Potts to the XY model continues for the Heisenberg model: The energy per site $E(T)/N$ is already well converged for a cooling process with a rather large $\gamma=0.001$, see Fig.~\ref{fig:heis_e}(a). At small $T$ the energy is well approximated by a straight line $E(T)/N=-1.5J_1+T$ corresponding to two quadratic modes per spin. Particularly, this result indicates the absence of local zero modes which are e.g. present in Heisenberg antiferromagnets with nearest neighbor couplings only (where the energy shows the well-known low-$T$ behavior $E(T)/N\sim11T/12$~\cite{chalker92}). The system's specific heat $c_v(T)$ in Fig.~\ref{fig:heis_e}(b) exhibits a peak at $T\approx0.14J_1$ which is again significantly lower compared to the three-state Potts and XY models. As we will discuss below, this peak again marks a crossover into a low temperature regime where thermal fluctuations slow down, however, the equilibration process is still qualitatively different compared to the three-state Potts and XY models.

An important property explaining these differences is that local low-energy defect states are unstable in the Heisenberg model. For the isolated fractons constructed in Sec.~\ref{sec:ground_heisenberg} this has already been discussed: There are paths in configuration space where a defect state can be continuously transformed into a ground state and along which the energy decreases monotonically. Vortex states are, likewise, unstable since local three-component spins do not support a topologically protected vorticity. Nevertheless, the relaxation process of defects is very slow, such that it can be numerically costly to obtain perfectly thermalized ensembles within classical Monte Carlo. Energetically our numerical outcomes seem to be close to thermal equilibrium [see Fig.~\ref{fig:heis_e}(a)], however, the real-space spin configurations still show remnants of decaying fractons.

An example for a spin configuration obtained at $L=100$, $T=0.01J_1$ and for the slowest simulated cooling rate $\gamma=0.001$ is shown in Fig.~\ref{fig:heis_fracton_example}(a). For each lattice site the angle between the spin chiralities [see Eq.~(\ref{chirality})] for the two adjacent $J_1$-triangles is plotted, i.e., for a homogeneous ${\mathbf q}=0$ state (domain wall in the three-state Potts model) this angle is zero ($\pi$). In the case of three-component Heisenberg spins, however, any intermediate value may also be assumed. As can be seen in Fig.~\ref{fig:heis_fracton_example}(a), the system features a network of domain walls where intersections correspond to fracton-like defects. A typical example is highlighted in the upper right corner (black circle) where two domain walls cross. However, the fading of two legs with increasing distance from the defect core indicates that this defect is in the process of decaying. To compare this spin arrangement with an intact fracton defined by a single defect hexagon, we plot in Fig.~\ref{fig:heis_fracton_example}(b) a state with $u=1$, $v=0.57$, and $w=-0.8$ as has been constructed in Sec.~\ref{sec:ground_heisenberg}. We generally find that for smaller cooling rates $\gamma$ these residual defect patterns become rarer such that we attribute them to incomplete thermalization. 

\begin{figure}
\includegraphics[width=0.9\linewidth]{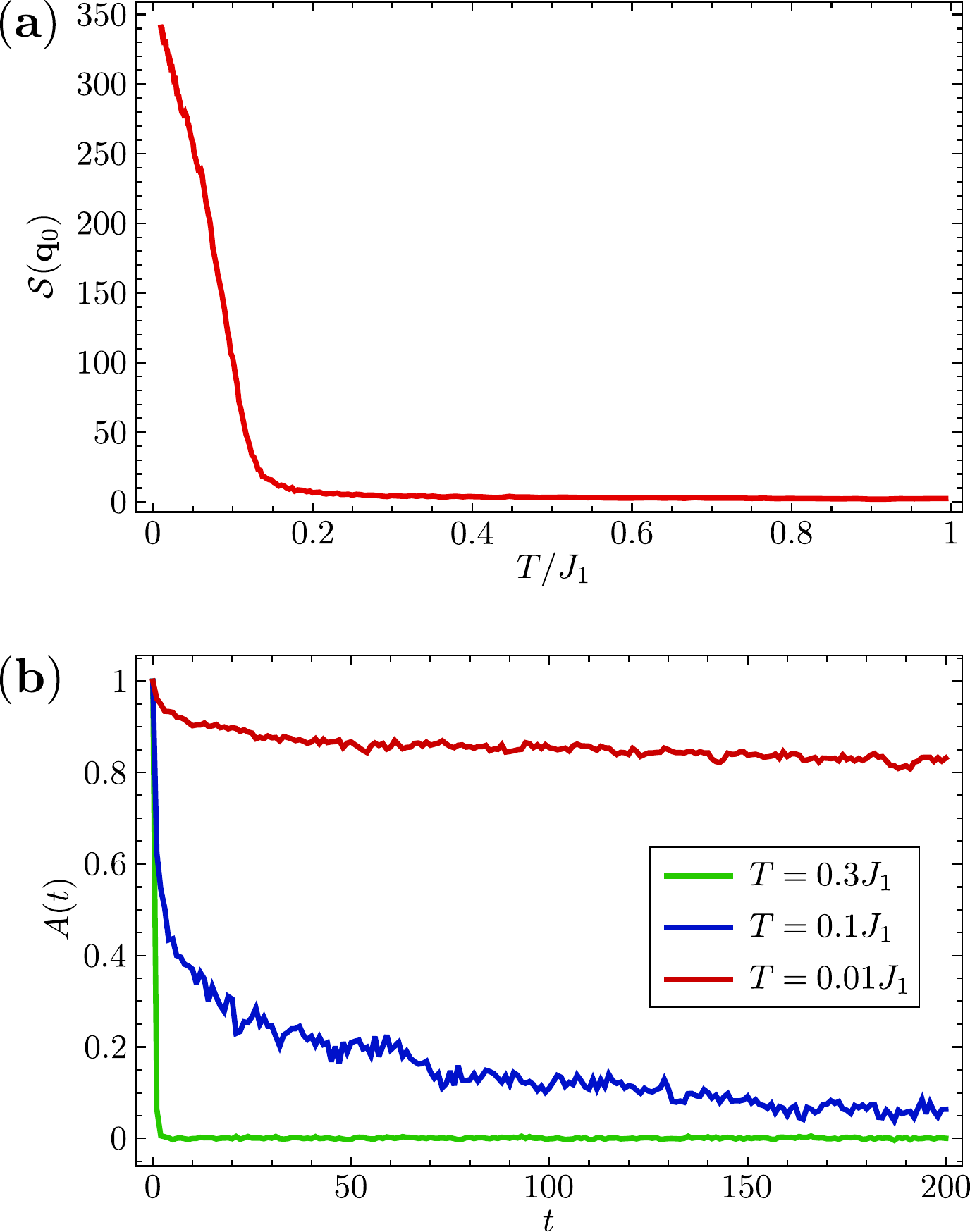}
\caption{   
\label{fig:peak_height_HB}(a) Spin structure factor $\mathcal{S}({\mathbf q}_0)$ of the Heisenberg $J_1$-$J_{5\bigtriangleup}$ model at the ${\mathbf q}=0$ order position ${\mathbf q}_0=(0,\pi/\sqrt{3})$ as a function of temperature for $L=100$ and $\gamma=0.001$. (b) Autocorrelation function of the Heisenberg $J_1$-$J_{5\bigtriangleup}$ model at different temperatures $T$ and $L=100$. The initial equilibration has been performed with a cooling rate of $\gamma=0.001$.}
\end{figure}
\subsubsection{Spin structure factor and autocorrelation function}
To further characterize the low-temperature regime we discuss the magnitude of the spin structure factor at the relevant ${\mathbf q}=0$ momenta which indicates the size of contiguous ${\mathbf q=0}$ order domains. As shown in Fig.~\ref{fig:peak_height_HB}(a), similar to the previous two models, $\mathcal{S}({\mathbf q}_0)$ exhibits a sharp increase at a temperature that matches the peak position in the specific heat. However, as a clear distinguishing feature $\mathcal{S}({\mathbf q}_0)$ does not show a plateau at small $T$ but instead keeps growing and is eventually only limited by the system's finite size. In accordance with the real-space plot in Fig.~\ref{fig:heis_fracton_example}(a) these results indicate that at low $T$ the system exhibits approximate ${\mathbf q}=0$ order configurations, however, the system does not freeze in such states but keeps slowly evolving towards more accurate ${\mathbf q}=0$ order realizations as the temperature is further lowered.

The autocorrelation function $A(t)$ in Fig.~\ref{fig:peak_height_HB}(b) reveals the same behavior. It is worth highlighting that in these results we have again eliminated effects of a global drift between the two times $t_0$ and $t_0+t$. Particularly, we have globally rotated the system at time $t_0+t$ such that ${\mathbf S}_i(t_0+t)$ at a given site $i$ points in the same direction as ${\mathbf S}_i(t_0)$. Additionally, a second overall rotation around ${\mathbf S}_i(t_0)$ needs to be performed which ensures that ${\mathbf S}_i(t_0)$, ${\mathbf S}_j(t_0)$, and ${\mathbf S}_j(t_0+t)$ all lie in the same plane (where $i$ and $j$ are arbitrary but fixed)~\cite{bilitewski19}. The autocorrelation function obtained this way again shows the two temperature regimes. While at temperatures above the heat-capacity peak a few Monte-Carlo steps are sufficient to completely suppress $A(t)$, in the low temperature regime a slow decrease is observed. However, even at $T=0.01J_1$ the autocorrelation function does not seem to saturate at a finite value for large $t$, consistent with an ongoing but slow thermalization process.  

In summary, both results in Fig.~\ref{fig:peak_height_HB} indicate a low-temperature regime with slow spin dynamics. In contrast to the previous two systems, however, this behavior is not caused by energy barriers and local energy minima in which the system may get trapped. Rather, the slow thermalization process stems from the equilibration dynamics of defects whose decay requires the simultaneous modification of an extensive number of spins.

\section{Conclusion}\label{sec:conclusion}

Fracton states of matter represent a vast landscape, stretching from the field  of quantum information and quantum many-body theory to experimental realizations.

In this work, we studied in detail how to realize fracton states on the kagome lattice, a paradigm of two-dimensional frustrated magnetism.
We analyzed an array of models with different interactions and different elementary degrees of freedom,
using a combination of analytical and numerical techniques.
They all have the characteristic subsystem symmetries and host fracton excitations, but their quantitative properties vary
depending on the model.
For example, 
the three-state Potts model
and the XY model
share the same ground state degeneracy structure aside from the global rotational symmetry
of the XY model;
however, 
the Heisenberg model
enjoys a much larger degeneracy
from subsystem operations on non-parallel lines.
The three-state Potts model
and the XY model 
also share similar fracton excitations that are stable and have a finite size,
while fractons in the Heisenberg model can smoothly decay into ground states at a power-law speed.

Using classical Monte-Carlo,
we studied the thermal properties of the 
models via their heat capacities, spin structure factors, and real space spin configurations.
In the three-state Potts model,
we discovered a crossover from a high-temperature paramagnetic phase to a low-temperature spin glass phase.
A similar crossover occurs in the XY model, where we additionally observe fractional 
vortices which, however, do not undergo a  Kosterlitz-Thouless transition  
due to their fracton nature.
While the  
Heisenberg model likewise shows a crossover into a low-temperature regime, its dynamics is not completely frozen,
due to the long-time instability of fractons.

We also find an unusual
deviation from the conventional
type-I fractons: In the kagome model, a single fracton cannot be isolated from a ground state by extending a fracton quadrupole to infinity. 
This, and various other low-energy properties of the model can be explained by viewing the kagome model as an embedding in the cubic fracton model.

Many interesting questions follow from this work. 
Our discovery suggests that
there is an underlying effective theory of fractonalized vortices
that is equivalent to the classical 2D fracton model with subsystem symmetries.
It will be interesting to pursue a clearer understanding of this equivalency.
Another useful development would be to extend the model to 
three dimensions.
For example, frustrated magnets on the pyrochlore lattice 
may be further good candidates for realizing fractons.
It is also interesting to upgrade the models 
to its quantum version, 
and explore the existence of fractons therein.

We hope that this work lays some groundwork for finding
experimentally  realistic scenarios to realize fractons,
and will be useful for both future experiments as well as
other model building efforts.

\section*{Acknowledgements}
We thank S. Parameswaran, S. Rachel,  Y. Iqbal, and Nic Shannon for fruitful discussions. We, particularly, acknowledge the valuable work of M. R\"o\ss ner which partially motivated this project. This work was partially supported by the German Research Foundation within the CRC183 (project A04).
HY is supported by the Theory of Quantum Matter Unit at Okianwa Institute of Science and Technology.

%


\end{document}